\RequirePackage{fix-cm}
\documentclass[smallextended]{svjour3}       % onecolumn (second format)
\smartqed  % flush right qed marks, e.g. at end of proof
\usepackage{graphicx}
\usepackage{color}
\usepackage{upgreek}

\usepackage{amssymb,natbib,verbatim}
\newcommand{\araa}{Annual Review of Astronomy and Astrophysics}
\newcommand{\apjl}{The Astrophysical Journal Letters}
\newcommand{\apj}{The Astrophysical Journal}
\newcommand{\apjs}{The Astrophysical Journal Supplement}
\newcommand{\aj}{The Astronomical Journal}
\newcommand{\ssr}{Space Science Reviews}
\newcommand{\aap}{Astronomy and Astrophysics}

\newcommand{\mnras}{Monthly Notices of the Royal Astronomical Society}

\newcommand{\nat}{Nature}
\newcommand{\bain}{Bulletin of the Astronomical Institutes of the Netherlands}
\newcommand{\jcap}{Journal of Cosmology and Astroparticle Physics}
\newcommand{\physrep}{Physics Reports}
\newcommand{\procspie}{Proc. SPIE}
\newcommand{\prd}{Physical Review D}

\newcommand{\lya}{Ly$\alpha$}
\newcommand{\ha}{H$\alpha$}
\newcommand{\arcmin}{\mbox{\ensuremath{^\prime}}}%    % fractional arcminute symbol: 0.'0
\journalname{Astronomy \& Astrophysics Review}
\begin{document}

\title{The realm of the galaxy protoclusters\thanks{This project received support from CNPq (308192/2013-3; 459040/2014-6;  400738/2014-7), FAPERJ (111.404/2014; 202.876/2015), and the Visiting Scholar Program of the Research Coordination Committee of the National Astronomical Observatory of Japan (NAOJ).}%Grants or other notes
%about the article that should go on the front page should be
%placed here. General acknowledgments should be placed at the end of the article.}
}
\subtitle{a review}

%\titlerunning{Short form of title}        % if too long for running head

\author{Roderik A. Overzier}

%\authorrunning{Short form of author list} % if too long for running head

\institute{R. A. Overzier \at
              Observat\'orio Nacional\\
              \emph{Rua General Jos{\'e} Cristino, 77; S{\~a}o Crist{\'o}v{\~a}o, Rio de Janeiro, RJ; CEP 20921-400; Brazil}\\%Tel.: +55-21-12345\\
              %Fax: +55-45-678910\\
              \email{overzier@on.br}           %  \\
            %of F. Author  %  if needed
}

\date{Received: date / Accepted: date}
% The correct dates will be entered by the editor

\maketitle

\begin{abstract}
The study of galaxy protoclusters is beginning to fill in unknown details of the important phase of the assembly of clusters and cluster galaxies. This review describes the current status of this field and highlights promising recent findings related to galaxy formation in the densest regions of the early universe. We discuss the main search techniques and the characteristic properties of protoclusters in observations and simulations, and show that protoclusters will have present-day masses similar to galaxy clusters when fully collapsed. We discuss the physical properties of galaxies in protoclusters, including (proto-)brightest cluster galaxies, and the forming red sequence. We highlight the fact that the most massive halos at high redshift are found in protoclusters, making these objects uniquely suited for testing important recent models of galaxy formation. We show that galaxies in protoclusters should be among the first galaxies at high redshift making the transition from a gas cooling regime dominated by cold streams to a regime dominated by hot intracluster gas, which could be tested observationally. We also discuss the possible connections between protoclusters and radio galaxies, quasars, and Ly$\alpha$ blobs. Because of their early formation, large spatial sizes and high total star-formation rates, protoclusters have also likely played a crucial role during the epoch of reionization, which can be tested with future experiments that will map the neutral and ionized cosmic web. Lastly, we review a number of promising observational projects that are expected to make significant impact in this growing, exciting field.
% \PACS{PACS code1 \and PACS code2 \and more}
% \subclass{MSC code1 \and MSC code2 \and more}
\end{abstract}

%\clearpage

%\tableofcontents

%\clearpage

\section{Introduction}

\subsection{Background}

Galaxy clusters play an important role in numerous aspects of extragalactic astrophysics and cosmology. Because of their extreme nature, galaxy clusters and their high redshift progenitor structures featured prominently in many early discussions of cosmological models (``steady-state'' versus ``Big Bang''), structure formation scenarios (``top-down'' versus ``bottom-up''), and the properties of the dark matter \citep[``hot'' versus ``cold''; e.g.,][]{zwicky39,sunyaev72,press74,white78,efstathiou81,davis85}. 
The important problem of galaxy cluster formation in which small gravitational instabilities grow by many orders of magnitude in an expanding universe, first increasing in size, then contracting and finally virializing in accordance with the scaling relations of clusters also paved the way for some of the first analytical and numerical calculations of structure formation \citep{vanalbada60,vanalbada61,aarseth63,peebles70,icke73}. In extragalactic astrophysics, clusters are key to tracing the formation of the most massive dark matter halos, galaxies and supermassive black holes (SMBHs) \citep[e.g., ][]{springel05}. They are essential for galaxy formation models because of the distinct properties of galaxies in dense environments. It has been shown that in order to correctly interpret the correlations observed between galaxy properties and environment, it is important, though challenging, to separate environmental effects operating in clusters from halo assembly bias \citep[``nurture'' versus ``nature''; see][for a review]{delucia07_review}. The latter suggests that the properties of galaxies in a halo may depend not only on the mass of that halo, but also on its formation time, and its effects may be particularly pronounced in massive halos \citep[e.g.,][]{gao07,cooper10,wang13,hearin15,lin16,miyatake16}. 
\begin{figure}[t]
\begin{center}
\includegraphics[width=0.7\textwidth]{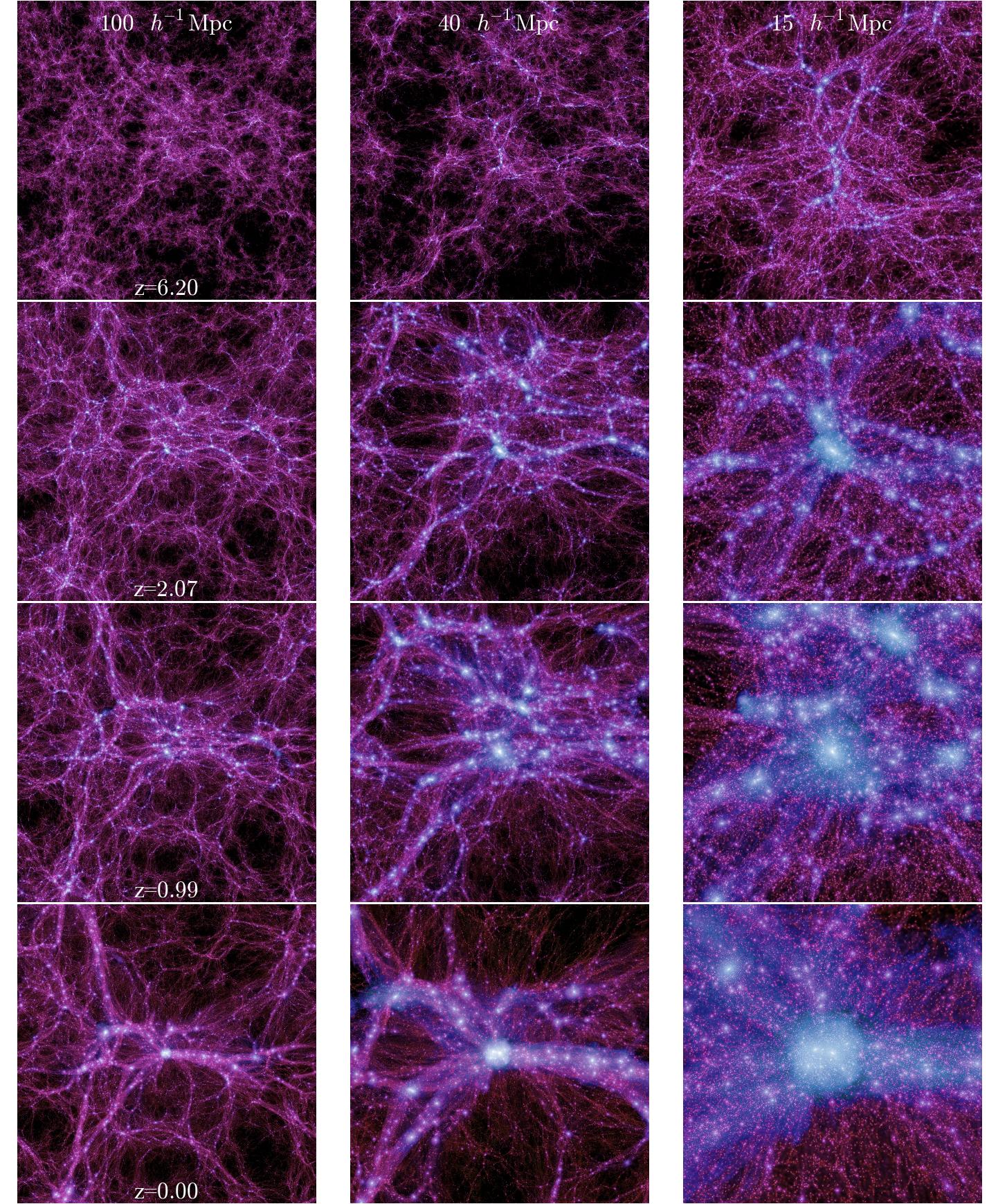}
\end{center}
\caption{\label{fig:boylankolchin09}The development of a massive Coma-like galaxy cluster in the Millennium II dark matter simulation. Panels show the dark matter distributions near the (proto-)cluster on 3 different scales (from {\it left-to-right} 
100, 40 and 15 $h^{-1}$ Mpc) and at 4 different epochs (from {\it top-to-bottom} $z=6.20$, $z=2.07$, $z=0.99$, and $z=0$). Prior to $z\sim2$ the main characteristic of the ``protocluster'' is its large-scale overdensity of dark matter (and associated gas and galaxies) that can be seen as early as $z\sim6$ on scales of several tens of $h^{-1}$ Mpc ({\it top row of panels}). 
Figure reproduced from Fig. 2 in \citet{boylankolchin09}.}
\end{figure}

The gravitational collapse of galaxy clusters is best studied theoretically in numerical cosmological simulations that trace the evolution of the large-scale structure as a function of time in three dimensions. There is a vast literature on simulations of galaxy clusters that focuses primarily on clusters as a tool for constraining cosmological parameters (e.g., by comparing the number abundance, masses and density profiles of clusters at different redshifts), and on understanding the complex baryonic processes associated with massive halos. However, the very early stages of galaxy cluster formation, which takes place mostly in the linear regime prior to virialization, are not very well known. This growth phase is marked by the evolution of numerous individual galaxies forming in rather loose aggregations of dark matter halos and intergalactic gas associated with extended, overdense regions of the cosmic web at high redshift. In Fig. \ref{fig:boylankolchin09} we show an example of a galaxy (proto-)cluster in a dark matter simulation \citep{boylankolchin09}. The different panels show the dark matter density distribution at one location at four different redshifts ($z\sim6,2,1,0$) and on three different scales (100, 40, 15 $h^{-1}$ Mpc). A number of key characteristics of the cluster formation scenario in a $\Lambda$CDM universe emerge from simply looking at these pictures. The general region of a forming cluster can already be identified as early as $z\sim6$ by means of a modest density contrast in the dark matter distribution seen on scales of several tens to a hundred Mpc. The initial matter distribution at high redshift consists of numerous smaller haloes that are clustered stronger than the dark matter, without a clear central halo, at least at $z\gtrsim2$. This is the protocluster stage of the collapse. At later stages, a strong central mass concentration builds up, and the surrounding structure becomes more and more filamentary with large voids. At $z\sim1$, several equally sized halos can be seen that will still merge to form the single final cluster. Many of these  larger halos prior to $z=0$ are quite elongated, while the final cluster at $z=0$ is spherical. At the lower redshifts ($z\lesssim1$) the large-scale structure surrounding the cluster appears to stabilize due to the increased expansion rate \citep{vikhlinin14}.

Figure \ref{fig:boylankolchin09} indicates that protoclusters can, in principle, be identified by their dark matter overdensities from very early on ($z\sim6$) provided that suitable baryonic tracers exist. Their properties are important for obtaining a full census of galaxy evolution in the universe. The $\sim$4 Gyr epoch between $z\sim4$ and $z\sim1$ was critical for the assembly of massive galaxies. During this epoch, both the star formation and quasar (QSO) activity peaked, and most of the stellar mass in massive galaxies was assembled into individual galaxies \citep[][]{madau14}. Observing this crucial epoch can give very important clues to the process of galaxy formation, especially in the densest regions: these (forming) group and cluster environments contain large quantities of co-eval galaxies seen at a time when most of the ``action'' is taking place, magnifying details of the assembly process, including the infall of matter from the filamentary cosmic web, the fueling of star formation, interactions between galaxies, co-evolution of galaxies and their SMBHs, formation of brightest cluster galaxies (BCGs), the first stages of heating and metal-enrichment of the intracluster medium (ICM), and the build-up of the intracluster light (ICL). The highest star formation rate (SFR) occurred in the high redshift progenitors of today's clusters, and those halos reached their peak SFRs at an earlier redshift compared to halos of lower mass \citep{behroozi13}. Thus, if we do not find and study these rare, massive structures at high redshift, we will lack vital empirical evidence necessary to understand cluster formation. Although some galaxy clusters with a prominent red sequence and a thermalized ICM have been identified up to $z\approx2$, detecting their higher redshift progenitors poses serious challenges because of the subtle density contrasts involved as well as their rarity. In a sense, the study of galaxy cluster evolution significantly lags that of galaxy evolution, for which the observational frontier currently lies at $z\approx10$ with large samples of galaxies available for statistical analysis at $z\lesssim6$ \citep{bouwens15a}. Just as galaxy evolution relies on connecting present-day galaxies with their higher redshift progenitors, it is important that we study the assembly of clusters to ever more distant progenitor objects up until the smallest density fluctuations from which they originated. 

\subsection{Scope and outline of this review}
\label{sec:scope}

The study of galaxy clusters goes back almost a century, and there exist many excellent reviews on galaxy clusters \citep[e.g.,][]{bahcall77,dressler84,norman07,allen11,rosati02,kravtsov12,vikhlinin14,bykov15}. These reviews focus on the many aspects of galaxy clusters, including their use as cosmological probes, dark matter (sub-)structure, BCGs and AGN feedback, red sequence evolution, environmental effects on galaxies, ICM enrichment and hydrodynamical evolution. A thorough review of galaxy clusters is beyond the scope of this paper, and we refer the reader to the aforementioned reviews and references therein. 
The (protocluster) environments of high redshift radio galaxies, known to be good tracers of high-density regions, were reviewed as part of \citet{miley08}, and many other recent papers referenced in this review present detailed views of many aspects of protoclusters. In this paper, we compile an introductory level review of this relatively young but rapidly growing field. The structure of this review is as follows. In Sect. \ref{sec:definitions} we will give some necessary definitions related to the term ``protocluster''. We will start by giving a brief overview of recent high redshift cluster surveys (Sect. \ref{sec:higzclusters}), followed by a description of the main techniques used for finding protoclusters (Sect. \ref{sec:protoclustersearch}). In Sect. \ref{sec:zoo} we present the existing sample of protoclusters selected from the literature and discuss the main observational characteristics such as redshifts, overdensities, and derived masses. The properties of protoclusters are described in more detail in Sect. \ref{sec:properties}, with an emphasis on the dark matter and gas accretion history of protoclusters (Sect. \ref{sec:evolution}), development of the cluster red sequence (Sect. \ref{sec:redsequence}), BCGs (Sect. \ref{sec:bcg}), and the comparison between galaxies in protoclusters and in the field (Sect. \ref{sec:environmentaleffects}). Next, we will take a closer look at interesting possible connections between protoclusters and Ly$\alpha$ blobs (Sect. \ref{sec:labs}) and radio galaxies and quasars (Sect. \ref{sec:rgqso}). We will discuss the role of protoclusters during the epoch of reionization in Sect. \ref{sec:reionization}. We conclude this review with a brief overview of new surveys and instruments that may have a high impact on this emerging field (Sect. \ref{sec:surveys}), and by highlighting a number of key open topics (Sect. \ref{sec:finalremarks}). 

\subsection{Definition of clusters and protoclusters}
\label{sec:definitions}

The term ``cluster'' is used differently by different authors. Many authors reserve this term for the most massive, virialized systems ($M\gtrsim10^{14}$ $M_\odot$), while referring to systems with masses between $\sim10^{13}$ $M_\odot$ and $\sim10^{14}$ $M_\odot$ as ``groups''. However, for lack of a real physical distinction between the two, other studies make no distinction and refer to all massive systems above some threshold mass collectively as ``groups'' or ``clusters''. In this review, we will follow the popular convention that clusters are virialized objects with total masses $\gtrsim10^{14}$ $M_\odot$ \citep[e.g., see][]{bower04}. Because the halo mass function is very steep at high masses, the  distinction between groups ($10^{13}<M/M_\odot<10^{14}$) and clusters ($M/M_\odot\ge10^{14}$) helps to more easily track their evolution when comparing to theoretical models or cosmological simulations. 

The study of protoclusters has been somewhat plagued by the fact that the term ``protocluster'' is also used differently by different authors and different studies. However, if we adopt the simple definition of a ``cluster'' given above, the meaning of the term ``protocluster'' becomes straightforward: a protocluster is a structure that will, at some stage, collapse into a galaxy cluster (i.e., a virialized object of $\ge10^{14}$ $M_\odot$ at $z\ge0$). This definition has the convenient implication that the combined space number density of clusters and protoclusters at any redshift is equal to the abundance of clusters today\footnote{Despite the working definition of a protocluster that we will adopt here, there are many alternative views of what constitutes a protocluster that are, at the least, just as valuable. For example, a very conservative approach would be to call a massive or overdense high redshift structure a cluster only when it meets a minimum set of conditions typical for clusters (e.g., detection of a thermal ICM in the X-rays or a well-defined cluster red sequence), and anything else a protocluster. Other definitions could be constructed based on the redshift at which half of the present-day mass of a cluster was assembled ($R_{1/2}$) or on the redshift where environmental effects from a dense gaseous medium or galaxy interactions begin to alter significantly the properties of the infalling and orbiting cluster galaxies (e.g., ram-pressure stripping, tidal stripping, dynamical friction, and quenching).}.  

It is important to keep in mind that even when we provide a strict definition of what constitutes a protocluster, it is really only practical in theory and simulations, and not in observations. The observational data are often insufficient to decide with absolute certainty if we are dealing with the progenitor of a galaxy cluster. Moreover, in order to be able to classify an object as a protocluster, we require not only detailed knowledge about the distant object itself, but also about its descendant at the present-day. Especially the latter requirement presents a serious challenge that typically can only be addressed in a statistical way rather than on an object-by-object basis. Note that for the lower redshift galaxy clusters this is typically not a problem, as the presence of the cluster red sequence, the velocity dispersion of the galaxies, or the X-ray, Sunyaev--Zel'dovich (SZ) effect or gravitational lensing mass profiles usually provide accurate enough constraints on the system's mass, and its dynamical and evolutionary state. 

In practice, many authors therefore use a somewhat stricter definition of the term protocluster as being a structure that is sufficiently overdense compared to its surroundings such that it can be recognized observationally. This does not conflict very strongly with the all-encompassing definition of all cluster progenitors that we use here. However, it is important to realize that for every protocluster observed, there are typically many other that have not yet developed a significant density contrast or that are missed in observations because of incompleteness or sensitivity. Another subtlety arises when we consider that many virialized clusters are still surrounded by other material that has yet to become part of the cluster through infall or merging \citep{chiang13,muldrew15}. In these cases, the protocluster could be seen as a much larger region that includes the entire structure. However, we will follow the current literature convention to call any collapsed object of at least $10^{14}$ $M_\odot$ a cluster, and call it a protocluster when such a massive core is not yet present. However, it is important to remember that cluster formation is a process that is occurring at all redshifts. An example of a protocluster at the relatively low redshift of $z\sim0.4$ is the super-group SG1120--1202 \citep{gonzalez05,kautsch08,smit15}, which is expected to collapse to form a Virgo-sized cluster by the present-day. In this review, however, we will focus on protoclusters at much higher redshifts. This choice is not entirely arbitrary as the redshift range $1.5\lesssim z\lesssim2.5$ covers the transition epoch for typical massive clusters and their protocluster progenitors.  

\section{Searching for protoclusters} 
\label{sec:observations}

\subsection{High redshift cluster surveys}
\label{sec:higzclusters}

There is a long and rich tradition of surveys aimed at detecting galaxy clusters at increasingly high redshifts. The main searches are centered around looking for concentrations of red sequence and other massive galaxies, or looking for the characteristic signature of hot cluster gas either in the X-rays or using the SZ effect. The frontier of cluster searches currently lies at $z\simeq1.5-2$, which has been identified as a very important epoch for massive galaxy clusters. 

The pioneering work on the Red Sequence Cluster Survey (RCS) by \citet{gladders00,gladders05}, and its successor, RCS-2 \citep{gilbank11}, showed that large samples of clusters can be efficiently selected out to $z\approx1$ by making use of standard optical filters combined with the $z$-band to pinpoint overdensities of galaxies as their Balmer or 4000 \AA\ break redshifts through the optical and NIR filters. By extending the red sequence search strategy to near- and mid-IR filters (e.g., using the colors $J-K$ or $z^\prime-[3.6~\upmu$m]), the red sequence technique has been successfully employed at $z>1$, for example using the UKIRT Infrared Deep Sky Survey \citep[UKIDDS;][]{andreon09} and the Spitzer Adaptation of the RCS \citep[SpARCS;][]{muzzin09,wilson09}. Although the $z^\prime-[3.6~\upmu$m] color nicely straddles the red sequence at these high redshifts, galaxies are about 2 magnitudes fainter at $z\sim2$ with respect to $z\sim1$, requiring imaging depths that pose a significant challenge to finding more distant clusters \citep{muzzin13}. One solution to this problem was obtained by \citet{papovich08}. To identify overdensities of co-eval galaxies, they used the redshifted 1.6 $\upmu$m `stellar bump' feature, which is ubiquitous in the spectra of most types of galaxies. At $z\gtrsim1$, the color [3.6 $\upmu$m] -- [4.5 $\upmu$m] increases monotonically with redshift up to $z\sim1.7$, after which it reaches a plateau. Foreground galaxies can furthermore be removed by involving an optical or near-IR filter \citep{papovich08,falder11,muzzin13}. In order to construct a high redshift sample of cluster candidates, \citet{papovich08} selected the $3\sigma$ overdensities from the density map of the stellar bump galaxies smoothed on a scale corresponding to the expected size of a cluster core. This resulted in a large sample of candidate (proto-)clusters with the number density and clustering expected of the progenitors of present-day clusters. One of the candidates was spectroscopically confirmed to be at $z=1.62$ \citep{papovich10}, while another $z=2.00$ cluster\footnote{We should note that several of these objects would be classified as protoclusters if we follow the definitions given in Sect. \ref{sec:definitions}. These objects have mass estimates based on either X-ray luminosity or velocity dispersion that formally lie below $10^{14}$ $M_\odot$ \citep[e.g., see][]{pierre12,gobat13,yuan14}.} having a strong red sequence of early-type galaxies was also identified using this technique \citep{gobat11,strazzullo13}. The Clusters Around Radio-Loud AGN (CARLA) survey targeted several hundred radio-loud quasars and radio galaxies at $z>1.3$ with the Spitzer Space Telescope, discovering several new mid-IR selected clusters \citep[][see also Sect. \ref{sec:rgqso2}]{galametz12,wylezalek13,cooke15a,cooke16,noirot16}. Using both the stellar bump method and the red sequence technique, \citet{muzzin13} have detected a rich cluster at $z=1.63$, while \citet{rettura14} found a large number of candidate clusters in the redshift range $1.3<z<2.0$ in the 100 deg$^{2}$ Spitzer South Pole Telescope Deep Field \citep[SSDF;][]{ashby13}.

Instead of color selections, that may rely on the type of stellar populations present in cluster galaxies as a function of redshift, a more complete selection of distant clusters can be made by searching for overdensities in redshift space, provided that accurate photometric redshifts over a large survey area are available \citep[e.g.,][]{vanbreukelen09,eisenhardt08,chiaberge10,spitler12,castignani14,chiang14,ascaso15,wang16}.  
\citet{strazzullo15} searched for overdensities of passive galaxies with photometric redshifts using nearest neighbor or fixed physical radius densities, finding several promising candidates in the range $1.5 < z < 2.5$. The IRAC Shallow Cluster Survey \citep[ISCS;][]{eisenhardt08} and its higher redshift extension, the IRAC Distant Cluster Survey \citep[IDCS;][]{stanford12}, searched for redshift overdensities to identify a large sample of $z>1$ cluster candidates of which many have since been confirmed, including the extremely massive cluster IDCS J1426+3508 at $z=1.75$ \citep[e.g.,][]{brodwin11,brodwin16,stanford12}. The cluster JKCS 041 at $z=1.80$ was discovered using the red sequence technique \citep{andreon09}. It has a large mass ($>10^{14}$ $M_\odot$) based on both the mass-richness relation and the detection of a hot ICM \citep{andreon14}. 

Focusing instead on their gaseous components as a means of discovering high redshift clusters, existing wide-field X-ray surveys are at the limits of their sensitivity for clusters at about $z\sim1.5$ \citep{rosati02,mullis05,stanford06}. Nonetheless, a recent deep 50 deg$^2$ cluster survey with {\it XMM-Newton} has led to the discovery of one of the highest redshift clusters, XLSSU J021744.1--034536 at $z\simeq1.91$, which was also detected using SZ effect follow-up observations \citep{mantz14}. SZ surveys are becoming more and more powerful for identifying high redshift clusters through the spectral distortion of the cosmic microwave background as its photons traverse the ICM. \citet{bleem15} found hundreds of new clusters in the 2500 deg$^2$ South Pole Telescope SZ survey (SPT-SZ). SPT-SZ clusters have a median redshift of $z\sim0.55$ with a significant tail extending to much higher redshifts, as evidenced by the discovery of SPT-CL J2040--4451 at $z=1.478$. Although the detection rate of very high redshift clusters using blind SZ effect and X-ray surveys is still lagging behind that of galaxy-based surveys, this situation may soon change due to upgraded SZ instrumentation at the SPT and the Atacama Cosmology Telescope, as well as new X-ray missions. 

\subsection{Protocluster searches at $z\gtrsim2$}
\label{sec:protoclustersearch}

As we move into the epoch during which the galaxies that are in (proto-)clusters become difficult to distinguish from galaxies that are in the field due to the absence of a well-defined red sequence or a hot ICM, the most conspicuous sign that indicates the presence of a (proto-)cluster is the relatively large concentration of galaxies and associated gas clustered in angular coordinates and redshift space. The main techniques for finding protoclusters are therefore constructed around this simple observational signature. Below we will review the main methods that are being used to find protoclusters.

\subsubsection{Surveys and ``serendipitous'' discoveries}

Because galaxy clusters are a fundamental component of the large-scale
structure of the present-day universe as they occupy the dense nodes
in the filamentary cosmic web, surveys capable of tracing the
formation of this cosmic web at high redshift are, in principle, also
surveys of cluster formation. In practice, the study of cluster
formation with standard galaxy surveys is extremely complicated
through a number of reasons. Clusters are rare objects in the local
universe, and extremely large cosmic volumes need to be surveyed to
find their progenitors. The density contrasts between (proto-)clusters
and the field are relatively small at high redshift, requiring
sensitive surveys and, ideally, good spectroscopic coverage in order
to confirm candidates and disentangle the real structures from the
projection effects. 

Several protoclusters have already been found as by-products of large
spectroscopic surveys. Good examples of such discoveries are the protoclusters at
$z=2.30$ and $z=3.09$ discovered in the HS1700+643 and SSA 22 fields,
respectively \citep{steidel98,steidel00,steidel05}. Both objects were found through rest-frame UV spectroscopy of candidate high redshift galaxies selected on the basis of the Lyman break. At present, the size of the cosmological volumes at high redshift that can be surveyed using this technique are still relatively small. However, it is possible to speed up this cumbersome process by pre-selecting the most overdense regions (in projection) from wide-field photometric surveys of, e.g., Lyman break galaxies or \lya\ emitters, and following up spectroscopically only the most overdense regions. Good examples of early structures hence identified are the large structures at $z=4.86$ and $z=6.01$ both in the Subaru Deep Field \citep[][]{shimasaku03,toshikawa12}, and at $z=5.7$ in the Subaru/XMM-Newton Deep Field \citep[][]{ouchi05}. Recently, \citet{toshikawa16} performed a protocluster survey in the Canada--France--Hawaii Telescope Legacy Survey (CFHTLS) deep fields covering an area of 4 deg$^2$ with sufficient filters and depth to make surface density maps of dropout galaxies at $z\simeq3-6$. They followed up spectroscopically only the most significant ($>4\sigma$) overdensities for which simulations predict that $\sim$80 \% are genuine protoclusters, finding several new systems. \citet{chiang14} used a catalog of accurate photometric redshifts in the 1.6 deg$^2$ COSMOS field to select overdense regions in three-dimensional space (sky position and redshift) that are good candidates of Coma-like protoclusters at $z\simeq2-3$. They found 36 candidates, several of which have since been confirmed to be genuine protoclusters \citep[e.g.,][]{chiang15,diener15,lee14b,mukae16,wang16}. The two-step approach of first identifying surface overdensities, possibly aided by photometric redshifts, in a wider field survey followed by spectroscopy is significantly faster than performing a spectroscopically complete survey, although it has the disadvantage that there is no complete spectroscopic sampling of the cosmic web and one is most sensitive to structures that are relatively compact in the sky plane. 

The VIMOS Ultra Deep Survey \citep[VUDS;][]{lefevre15} is currently the largest spectroscopic survey (in terms of spectra per square degree) capable of sampling the cosmic web at high redshift ($z\simeq2-6$) with a sampling that is sufficiently dense and uniform to uncover a wide range of structures including protoclusters. The survey has already resulted in two exceptionally overdense, massive structures at $z=2.9$ and $z=3.3$ \citep{cucciati14,lemaux14}, and several tens more are expected to follow after analysis of the full VUDS data set.   

A large number of protoclusters were discovered serendipitously using data from the Planck survey. The Planck all-sky (sub-)millimeter maps contain a large number of compact (unresolved) ``cold'' sources \citep{clements14,dole15}. A large fraction of these sources are believed to be overdensities of star-forming galaxies in overdense regions at $z\simeq2-4$ with significant far-infrared emission redshifted into the 353--857 GHz frequency range \citep{dole15}. The Planck selection resulted in over one thousand candidates over 26 \% of the sky, of which about two hundred were followed up with Herschel to obtain higher resolution data. \citet{flores-cacho15} showed that at least one of the structures is due to two overlapping high redshift structures along the line of sight, at $z\sim1.7$ and $z\sim2$. Although most of the Planck sources remain to be confirmed spectroscopically, the uniform selection method and the large survey area make this a highly promising new search technique. The Planck selection is also interesting because it complements other protocluster selections that are based on optical- and NIR-selected galaxies (relatively dust-free, star-forming galaxies and passive galaxies). 

\subsubsection{Biased tracer techniques}
\label{sec:btt}

Finding galaxy (proto-)clusters at high redshift is like searching for a needle in a haystack. 
The biased tracer technique attempts to bypass the complications of
carrying out a very deep and wide galaxy survey by directly targeting the immediate
environment of previously identified galaxies that are known to be good tracers of
massive forming systems. Early protocluster studies have therefore
targeted objects for which there is strong evidence that they are
progenitors of the very massive galaxies that populate the centers of
today's clusters. 

One such class is that of the high redshift radio galaxies that have many properties indicative of progenitors of local BCGs \citep[e.g.,][]{best98,zirm03,miley08,collet15}. 
A substantial number of protoclusters were found by targeting the environment of previously known radio galaxies \citep[e.g.,][]{pascarelle96,lefevre96,pentericci00,kurk00,kurk04a,venemans02,venemans04,venemans05,venemans07,chiaberge10,kuiper11b,hatch11a,hatch11b,mawatari12,hayashi12,wylezalek13,cooke14}. In many of these studies, the first step was to use a narrow-band filter to identify candidate emission line galaxies such as Ly$\alpha$ emitters (LAEs) or H$\alpha$ emitters (HAEs) near the redshift of the radio source, often followed up by spectroscopy in order to confirm the candidates. By comparing the surface or volume densities of the emission line candidates near the radio sources with those obtained from field surveys, a substantial number of radio sources at high redshift was found to be associated with overdensities that were interpreted as protoclusters. The overdense fields were subsequently followed up at other wavelengths, showing in many cases that the overdensities are accompanied by overdensities of other galaxy populations as well, including red (sequence) galaxies \citep[e.g.,][]{kurk04b,kodama07,zirm08}, HAEs \citep{kurk04a,hatch11b,shimakawa14,cooke14}, Lyman Break Galaxies (LBGs) \citep[e.g.,][]{intema06,overzier06,overzier08,capak11} or Sub-mm Galaxies (SMGs) \citep[e.g.,][]{debreuck04,dannerbauer14,rigby14}.

Like the radio galaxies, the population of QSOs at high redshift is also frequently associated with massive galaxy and galaxy cluster formation. Many studies have targeted QSOs at $z\simeq2-6$ in order to study their immediate environments and look for protoclusters, sometimes finding overdensities and sometimes finding no difference with the random expectation \citep[e.g., ][]{djorgovski03,wold03,kashikawa07,stevens10,falder11,matsudasmail11,trainor12,husband13,adams15,hennawi15}. Much of the observational work has concentrated on the environment of the highest redshift QSOs at $z\gtrsim6$ \citep[e.g.,][]{stiavelli05,kim09,overzier09a,banados13,morselli14,simpson14}, because simple theoretical arguments and some simulations suggest that these should be clear cases of highly overdense regions \citep[e.g.,][]{springel05}. This is the topic of Sect. \ref{sec:rgqso} of this review.

Other sources that have been suggested to be good tracers of overdense regions at high redshift are \lya\ blobs (LABs) and SMGs. LABs are believed to trace large reservoirs of cool intergalactic gas associated with dense locations in the cosmic web, where it is being ionized and illuminated by a powerful AGN or starbursts (see Sect. \ref{sec:labs}).  
Narrow-band observations of protoclusters have shown that these LABs are preferentially found in overdense regions at high redshift, motivating the search for protoclusters near LABs as well \citep{prescott08,prescott12}. Dusty star-forming galaxies are believed to be the progenitors of present-day massive elliptical galaxies. Overdensities of these galaxies at high redshift could thus indicate a protocluster seen at an early stage when a large number of proto-ellipticals were going through their main star formation phase. This has motivated many studies to target the dusty galaxy population in protoclusters \citep[e.g.,][]{stevens03,greve07,priddey08,daddi09,carrera11,ivison13,casey15,dole15,valtchanov13,dannerbauer14,rigby14,umehata14,umehata15}. 

\subsubsection{Gas absorption studies: a promising new technique}
\label{sec:igm}

A relatively new technique for the selection of protoclusters is to
make use of the fact that overdense regions in the early universe are
not only overdense in dark matter and galaxies, but should also
contain large quantities of cold or warm, dense gas that can
be detected in absorption against a suitable background
continuum source such as QSOs and galaxies. Several protoclusters have been detected through neutral hydrogen absorption along the line of sight to background quasars \citep[e.g.,][]{francis96,steidel98,hennawi15}. Other structures have been detected by stacking the spectra of background galaxies, and searching for \lya\ or Mg II $\lambda\lambda$2797,2803 absorption at the redshift of the overdensity \citep[e.g.,][]{giavalisco11,cucciati14}. 

Recent authors have therefore suggested to search for the presence of protoclusters through absorption in a more systematic way. By taking spectra of a background population (star-forming galaxies and QSOs) that is sufficiently densely spaced on the sky, it will be possible to perform ``tomography'' of the cosmic web simply by mapping out the \lya\ absorption caused by the neutral hydrogen in the IGM \citep{stark15,lee14a,cai15}. 
Initially, authors suggested to use sightlines to background QSOs drawn from large spectroscopic surveys \citep[e.g.,][]{caucci08}. Even when the sightlines to background objects are spaced relatively sparsely, a larger than average optical depth that is coherent on large scales along the line of sight can be used as a tracer for highly overdense regions. \citet{frye08} and \citet{matsudarichard10} identified a $\sim80$ Mpc overdense region through the imprint of many overlapping absorbing HI clouds seen in the spectrum of a galaxy at $z=4.9$. \citet{finley14} serendipitously found a number of highly clustered regions of absorbing gas at $z=2.7$ along the sightlines to two background QSOs separated by about 100 kpc on the sky. The absorbers extend over about 2000 km s$^{-1}$ or 6.4 $h_{70}^{-1}$ Mpc, and may be due either to a protocluster or a filament seen along the line of sight. \citet{cai15} used simulations to study the correlation between the optical depth of the IGM and mass on scales of (15 $h^{-1}$ Mpc)$^3$ (i.e., protocluster scale regions) at $z=2.5$, finding that the highest effective optical depth regions trace the most massive structures. Although some signal confusion is expected from individual Damped Lyman-$\alpha$ Absorbers (DLAs) and clustered Lyman Limit Systems (LLSs), they show that they can pinpoint several previously known and unknown overdensities by means of this technique. 
\begin{figure}[t]
\begin{center}
\includegraphics[width=0.7\textwidth]{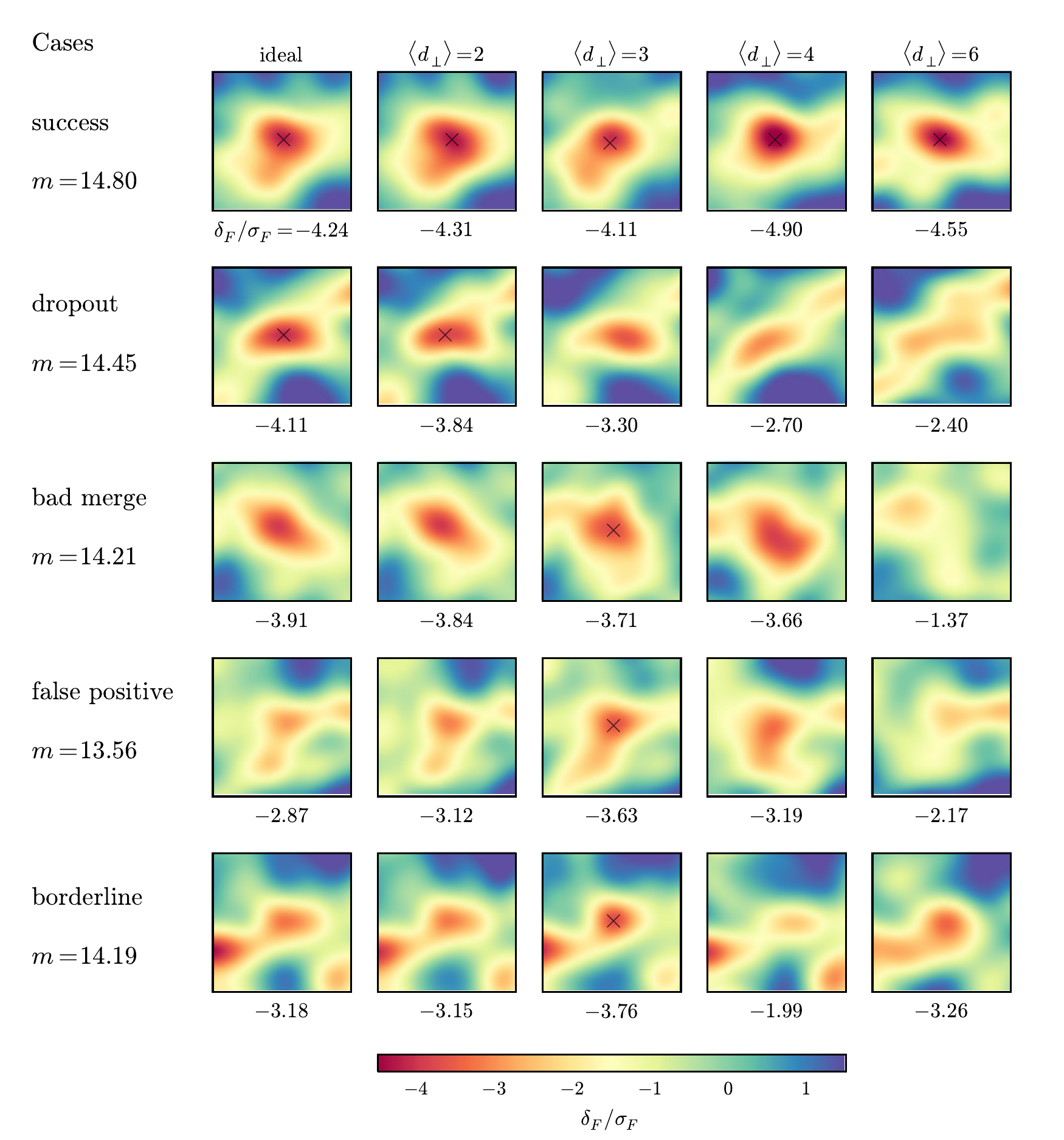}
\end{center}
\caption{\label{fig:stark15}Mock absorption maps for tomographic mapping of the IGM as predicted for 5 massive structures at $z=2.5$.  
Panels show the ideal and reconstructed maps for different average transverse separations of the background sources ranging from 2--6 Mpc ({\it left-to-right}), and for various scenarios \citep[{\it top-to-bottom}, see][for details]{stark15}. Figure reproduced from Fig. 8 in \citet{stark15}.}
\end{figure}

\begin{figure}[t]
\begin{center}
\includegraphics[width=\textwidth]{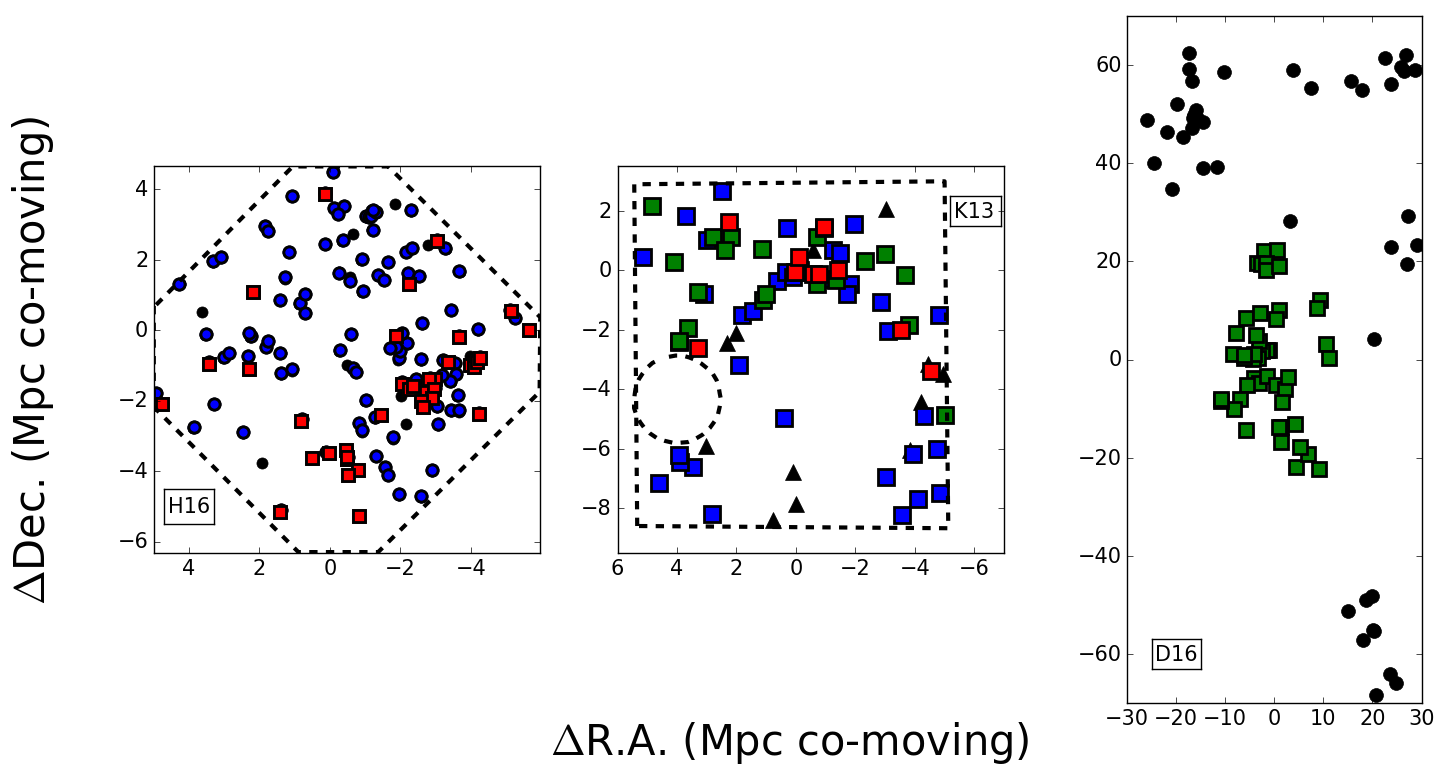}
\end{center}
\caption{\label{fig:zoo_xy}Examples of protoclusters, highlighting the spatial distributions of member galaxies in these overdense structures. The left panel shows the distribution of galaxies with photometric and spectroscopic redshifts in a protocluster at $z\approx1.62$ [{\it red squares} red galaxies satisfying $z-J>1.3$; {\it blue circles} galaxies satisfying SFR$>$5 $M_\odot$ yr$^{-1}$; {\it black circles} other galaxies; \citet{hatch16a})]. The middle panel shows a protocluster at $z\approx2.16$ ({\it red, green} and {\it blue squares} HAEs with relatively red, intermediate and blue colors; {\it black triangles} HAEs not detected in the $K$-band; \citet{koyama13a}). The right panel shows a $z=3.78$ protocluster in the Bootes field ({\it green squares} LAEs at $3.77<z<3.81$; {\it black circles} other LAEs; \citet{dey16}).}
\end{figure}

At magnitudes of $g\gtrsim24$ the number density of star-forming galaxies (LBGs) exceeds that of the QSOs by more than a factor of 10 \citep[e.g., see Figure 1 in][]{lee14a}, allowing tomographic IGM maps to be constructed with a much higher spatial resolution. Until recently, large spectroscopic surveys reaching these magnitudes over large cosmological volumes were unfeasible, but this situation is changing. The observational requirements for performing such a survey at $z\approx2$ were investigated by \citet{lee14a} and \citet{stark15}. \citet{stark15} looked at the specific absorption signatures of protoclusters using mock maps expected for a range of typical separations of background sightlines. A figure from that study is presented in Fig. \ref{fig:stark15}, which shows the mock absorption maps predicted for 5 massive structures at $z=2.5$ and the likelihood of identifying such structures correctly. \citet{lee14a} furthermore showed that with limiting $g$-band magnitudes of 24.0, 24.5, and 25.0 (corresponding to sightline densities of $\sim$360, 1200, and 3300 galaxies deg$^{-2}$), a three-dimensional map of the IGM can be reconstructed having a spatial resolution of only a few Mpc. They further show that the experiment could be performed with moderate resolution ($\simeq1000$) multiplexed spectrographs on 8--10m class telescopes covering an area of 1 deg$^2$ (volume of order $10^6$ $h^{-3}$ Mpc$^3$) in 100 h. A \lya\ tomographic survey is more efficient for mapping the three-dimensional cosmic web than a galaxy redshift survey at the same redshift. If large cosmological volumes can be mapped, the \lya\ tomographic mapping technique would be one of the most powerful probes for uncovering for the first time very large, unbiased samples of protoclusters. In order to demonstrate the technique on real data, \citet{lee14b} applied the \lya\ tomographic mapping technique in a $5'\times14'$ region of the COSMOS field. By targeting a small number of star-forming galaxies at $z>2.3$, a map of the large-scale structure as traced by \lya\ absorption was made over a $6\times14\times240$ Mpc region at a spatial resolution of $\approx3.5$ $h^{-1}$ Mpc. The map resolution was sufficient to show that galaxies follow several high-density peaks in the absorbing hydrogen distribution. Interestingly, the highest density feature in the absorption map of \citet{lee14b} lies along the line of sight toward a large-scale overdensity of LAEs at $z=2.44$ found independently by \citet{chiang15}, showing the potential for discovering protoclusters using the \lya\ tomography technique. 

\section{The protocluster ``zoo''}
\label{sec:zoo}

\subsection{Overview of objects found to date}

\begin{figure}[t]
\begin{center}
\includegraphics[width=\textwidth]{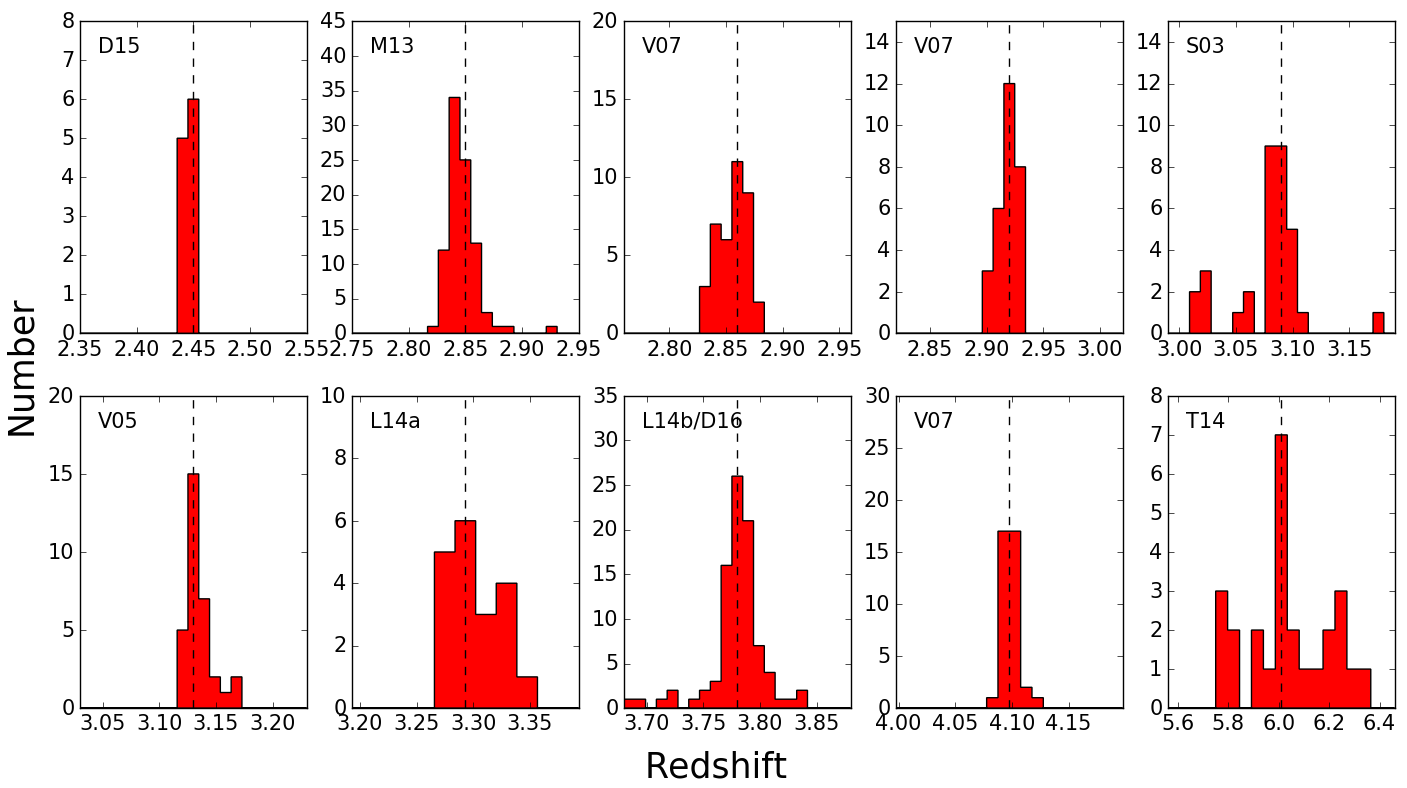}
\end{center}
\caption{\label{fig:zoo}Examples of protoclusters at $z\sim2-6$, highlighting the relatively narrow redshift distributions of spectroscopically confirmed galaxies in these overdense structures. References: \citet[][D15]{diener15}, \citet[][M13]{mostardi13}, \citet[][V07]{venemans07}, \citet[][S03]{steidel03}, \citet[][V05]{venemans05}, \citet[][L14a]{lemaux14}, \citet[][L14b]{kslee14}, \citet[][D16]{dey16} and \citet[][T14]{toshikawa14}.}
\end{figure}

\begin{figure}[t]
\begin{center}
\includegraphics[width=\textwidth]{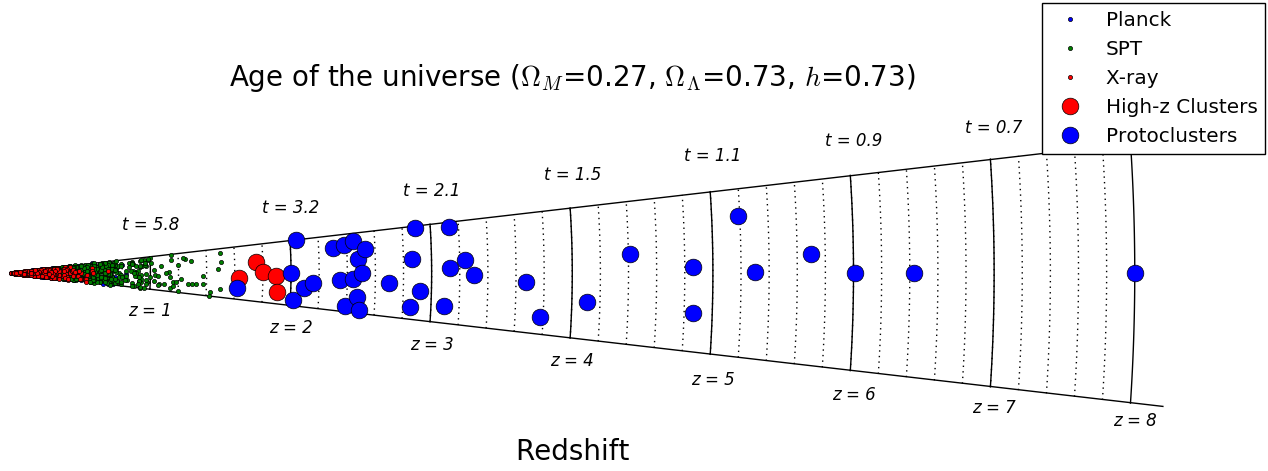}
\end{center}
\caption{\label{fig:pcredshifts}The distribution of redshifts of protoclusters selected from the literature ({\it large blue symbols}). The data for clusters below $z=1.5$ were taken from the compilation of clusters detected in X-ray and SZ surveys of \citet{bleem15}. Large red symbols are high redshift clusters at $z>1.6$. The position of objects along the polar axis holds no information and is used for visibility purposes only. Redshifts and ages (in Gyr) are indicated along the radial axes.}
\end{figure}

The techniques described in the previous section have resulted in a wide range of interesting structures that all highlight different aspects of cluster formation. However, taken together it is a highly heterogeneous data set, and it is currently still difficult to paint a complete picture of cluster evolution using these data. So what has been found so far? Some illustrative examples of the sky and redshift distributions of galaxies in protoclusters are shown in Figs. \ref{fig:zoo_xy} and \ref{fig:zoo}. These protoclusters were identified based on their large-scale (several to several tens of Mpc comoving) overdensities of various galaxy populations (e.g., star-forming galaxies, LBGs, HAEs and LAEs) that are strongly clustered in projected sky coordinates (Fig. \ref{fig:zoo_xy}) or in redshift space (Fig. \ref{fig:zoo}). In Fig. \ref{fig:pcredshifts} we present the redshift distribution of all protoclusters identified to date. We have based our compilation on the objects listed in Table 5 of \citet{chiang13}, supplemented by new objects discovered since \citep{cucciati14,lemaux14,kslee14,diener15,chiang15,casey15,toshikawa14,toshikawa16,ishigaki16,wang16}. We have included all structures at $z\gtrsim1.5$ having estimates of their present-day masses in excess of $\sim10^{14}$ $M_\odot$. About half of all known protoclusters are found in the $\sim1$ Gyr epoch between $z=2$ and $z=3$, while the other half is found in the $\sim1.5$ Gyr epoch between $z=3$ and $z=8$. Note, however, that we have not included here candidates resulting from ``snap-shot'' type and statistical surveys that have uncovered large numbers of as of yet unconfirmed candidates \citep[e.g.,][]{wylezalek13,dole15,chiang14,franck16}. The total number of potential protoclusters found is therefore probably substantially larger than that shown in Figure \ref{fig:pcredshifts}, as discussed elsewhere in this review. 

\begin{figure}[t]
\begin{center}
\includegraphics[width=\textwidth]{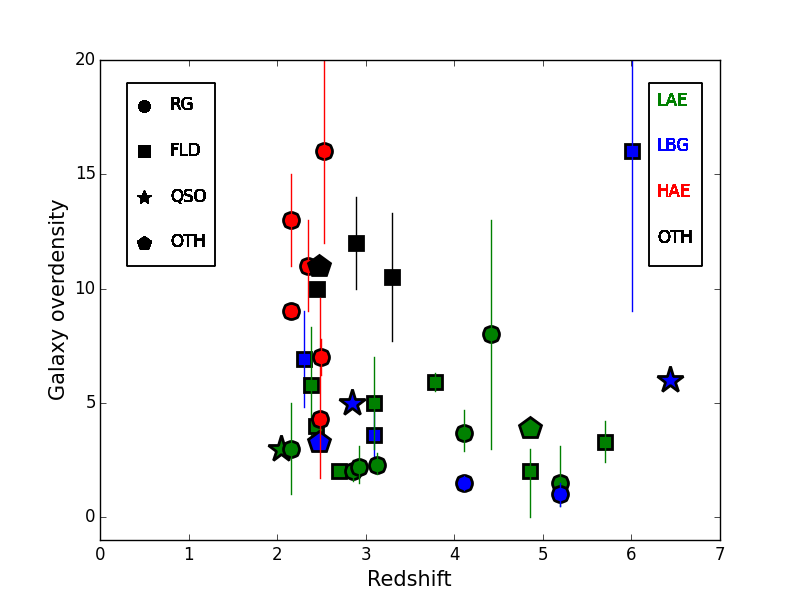}
\end{center}
\caption{\label{fig:pcoverdensities}The diversity of overdensities of galaxies measured for a selection of protoclusters at $z>2$. Symbols indicate protoclusters found using radio galaxies ({\it circles}), quasars ({\it stars}), other tracers ({\it pentagons}), and in the field ({\it squares}). Colors indicate overdensity values measured using LAEs ({\it green}), LBGs ({\it blue}), HAEs ({\it red}) or other types of galaxies ({\it black}). The measurements collected in this figure only serve to illustrate the range of overdensities typically associated with protoclusters, and should generally not be compared directly, either in the absolute or in the relative sense without taking into account the different selection and measurement techniques used by each study.}
\end{figure}

Figure \ref{fig:pcoverdensities} shows the range of galaxy overdensities ($\delta_\mathrm{gal}$) that have been measured for a subset of the protoclusters shown in Fig. \ref{fig:pcredshifts}. For each study, we have indicated the galaxy type that was used to measure the overdensities (LAEs, HAEs, LBGs or other), and whether the protocluster was found in the field or using a tracer (radio galaxy, QSO or other). These results show that although there already exists a relatively rich data set of all kinds of cluster progenitors, these data are rather heterogeneously selected and cover a wide redshift range. The measurements collected in this figure only serve to illustrate the range of overdensities typically associated with protoclusters, and should generally not be compared directly, either in the absolute or in the relative sense without taking into account the different selection and measurement techniques used by each study. Nonetheless, the range of overdensities measured for LAEs, LBGs and HAEs at these redshifts are broadly consistent with the overdensities expected for galaxy protoclusters based on the simulations of \citet{chiang13}, who calculated the overdensities of galaxies with SFRs $>$1 $M_\odot$ yr$^{-1}$ in a volume of (15 $h^{-1}$ Mpc)$^3$ around protoclusters.

\subsection{Why we believe these are protoclusters}
\label{sec:why}

\begin{figure}[t]
\begin{center}
\includegraphics[width=\textwidth]{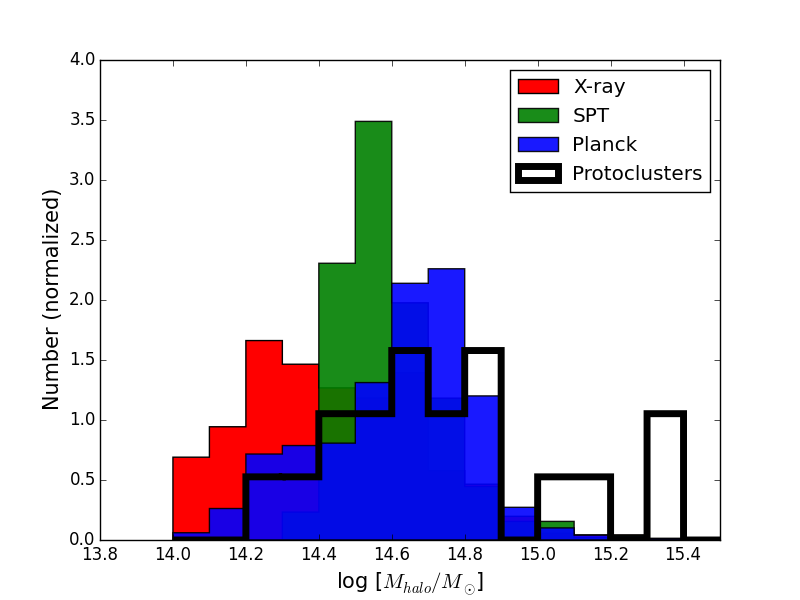}
\end{center}
\caption{\label{fig:pcmasses}The distribution of present-day ($z=0$) masses derived for the  protocluster sample from Fig. \ref{fig:pcredshifts} ({\it black histogram}), compared to the mass distribution of galaxy clusters discovered as part of X-ray ({\it red histogram}), South Pole Telescope (SPT, {\it green histogram}), and Planck ({\it blue histogram}) surveys \citep{bleem15}.}
\end{figure}

The structures shown in Figs. \ref{fig:zoo_xy}, \ref{fig:zoo}, \ref{fig:pcredshifts}, \ref{fig:pcoverdensities} span a rather large range in redshift, (co-moving) size, (projected) shape, and galaxy overdensity. In Fig. \ref{fig:pcmasses}, we show the distribution of the present-day masses estimated for the protoclusters selected from the literature. The median protocluster has $\mathrm{log}(M/M_\odot)_{z=0} = 14.6$, and the sample includes some exceptional protoclusters with ($z=0$) masses of $\gtrsim10^{15}$ $M_\odot$ \citep[e.g.,][]{venemans02,steidel05,cucciati14,lemaux14,diener15}.  These present-day masses of protoclusters span the full range of masses of typical galaxy clusters found in X-ray and SZ effect surveys \citep[see][for a recent compilation]{bleem15}, suggesting that the term ``protocluster'' is justified. 

How were these mass estimates obtained? In order to show that some overdense region of galaxies identified at high redshift will evolve into a present-day cluster, at least two requirements need to be fulfilled. First, the overdensity of the region needs to be large enough to enable it to detach from the expanding part of the universe around it, contract, and collapse. To guarantee that the region will be a galaxy cluster, the total mass of the final collapsed region needs to be at least $10^{14}$ $M_\odot$. Second, because we are interested in the progenitors of present-day galaxy clusters, this process must be completed before $z=0$. There exist several approaches that can be used to show that many of the structures found to date indeed fulfill these requirements, as discussed below. 

\subsubsection{Spherical collapse model}

\begin{figure}[t]
\begin{center}
\includegraphics[width=0.7\textwidth]{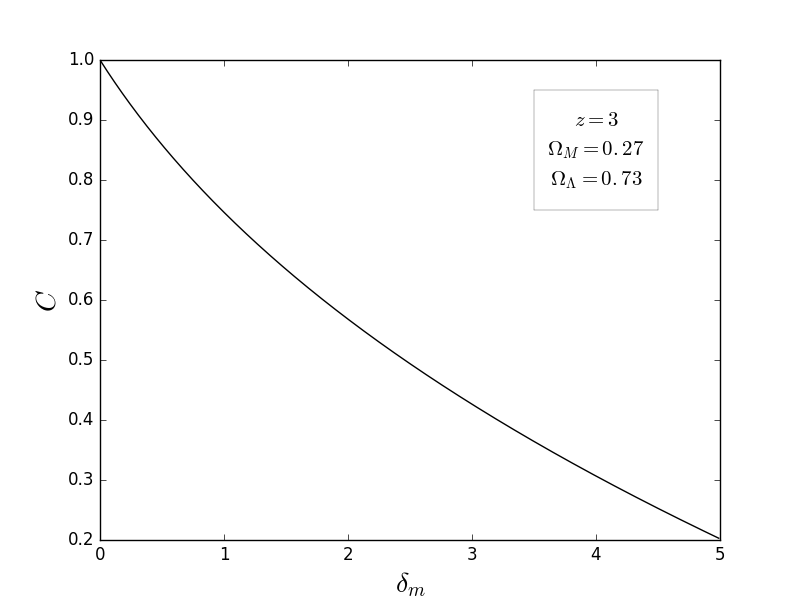}
\end{center}
\caption{\label{fig:c_delta_m}Redshift space distortion ($C$) as a function of the matter overdensity ($\delta_\mathrm{m}$) for $z=3$. $C$ gives the fractional compression of a protocluster volume in redshift space, and the fractional enhancement of the observed galaxy overdensity. See text for details.}
\end{figure}

Perhaps the simplest approach applied to protocluster data is to use the approximations of the spherical collapse model for a homogeneous sphere \citep{steidel98}. In this case, the total present-day mass of the protocluster is given by $M_{z=0}=\bar{\rho}V_\mathrm{true}(1+\delta_\mathrm{m})$, where $\bar{\rho}$ is the mean density of the universe, $\delta_\mathrm{m}$ is the dark matter mass overdensity, and $V_\mathrm{true}$ is the comoving volume of the protocluster in real space. The volume can be estimated from the transverse dimensions of the observed overdensity on the sky and its comoving depth corresponding to the width of the redshift distribution of the protocluster galaxies. A complication arises due to redshift space distortions, causing the overdense region to appear slightly compressed in redshift space, and the apparent overdensity to appear slightly enhanced from the true value \citep{kaiser87}. The true volume is given by $V_\mathrm{app}/C$, while the mass overdensity is related to the observed galaxy overdensity, $\delta_\mathrm{gal,obs}$, through $1+b\delta_\mathrm{m}=C(1+\delta_\mathrm{gal,obs})$, where $b$ is the bias parameter (defined as $\delta_\mathrm{gal}/\delta_\mathrm{m}$), and $C$ accounts for the redshift space distortion of the collapsing structure. The redshift space distortion itself naturally depends on the size of the mass overdensity, $C=1+f-f(1+\delta_\mathrm{m})^{1/3}$ \citep{steidel98}, where $f(z)=\Omega_M(z)^{0.6}$ for $\Lambda$CDM cosmologies \citep{linder05}. We are helped by the fact that the mass overdensities of protoclusters are typically still in the linear regime, with values for $C$ in the range of 0.4--1 for $\delta_\mathrm{m}$ between 0 and 3 (see Fig. \ref{fig:c_delta_m}). For flat $\Lambda$CDM models, $C$ furthermore only weakly depends on redshift. 

The spherical collapse mass overdensity can also be used to predict the future of the overdense region. By ``evolving'' the linear theory peak height, $\delta_\mathrm{L}$, forward in time, one can estimate at what redshift the linear overdensity will pass the critical density for collapse of a spherical symmetric perturbation \citep[$\delta_c\simeq1.69$;][]{weinberg13}. The linear overdensity $\delta_\mathrm{L}$ is directly related to $\delta_m$ and can be estimated using an analytic approximation or using the result from numerical simulations \citep[e.g.,][]{bernardeau94,mo96,jenkins01}. $\delta_\mathrm{L}$ evolves with redshift according to $\delta_\mathrm{L}(z_2)=D(z_1)\delta_\mathrm{L}(z_1)/D(z_2)$, with $D(z)\propto g(z)/(1+z)$ and $g(z)$ the cosmological growth factor \citep{carroll92}. Using this methodology, it has been shown that many of the protoclusters observed are indeed sufficiently massive and overdense that they will become genuine clusters \citep[e.g.,][]{steidel98,venemans02,steidel05,overzier08,cucciati14,lemaux14}.  

\subsubsection{Numerical simulations}

An alternative, and perhaps more powerful, method to interpret the protocluster data is to compare the observational findings directly to the output from numerical cosmological simulations of galaxy formation \citep[e.g.,][]{governato98,suwa06,delucia07,venemans07,overzier09a,angulo12,chiang13,chiang14,chiang15,sembolini14,granato15,muldrew15,contini15,miller15,hatch16a}. This technique has several advantages over the analytical approximations method described above. First, state-of-the-art cosmological simulations today probe cosmological volumes of $>0.1$ Gpc$^3$, i.e., large enough to provide good statistical information on the properties of clusters and their high redshift protocluster progenitors (typical space densities of $10^{-9}-10^{-6}$ Mpc$^{-3}$). Second, the spherical collapse model is not an accurate description of the formation of clusters (see Fig. \ref{fig:boylankolchin09}). Simulations naturally account for non-spherical collapse as well as non-linear gravitational effects. Third, in simulations (hydro, semi-analytical or abundance-matching based methods) that give predictions for the observable properties of galaxies, selection effects and other observational limitations can be taken into account. This means that we can much more accurately compare an overdensity of galaxies found at some redshift in an observational survey with similar-looking objects found in the simulations. An important caveat with these simulation-based methods, however, is that the results will depend on the accuracy of the simulation. If the cosmology or the physics of galaxies and clusters in the simulation differ  significantly, one may misinterpret the data.

\begin{figure}[t]
\begin{center}
\includegraphics[width=0.45\textwidth]{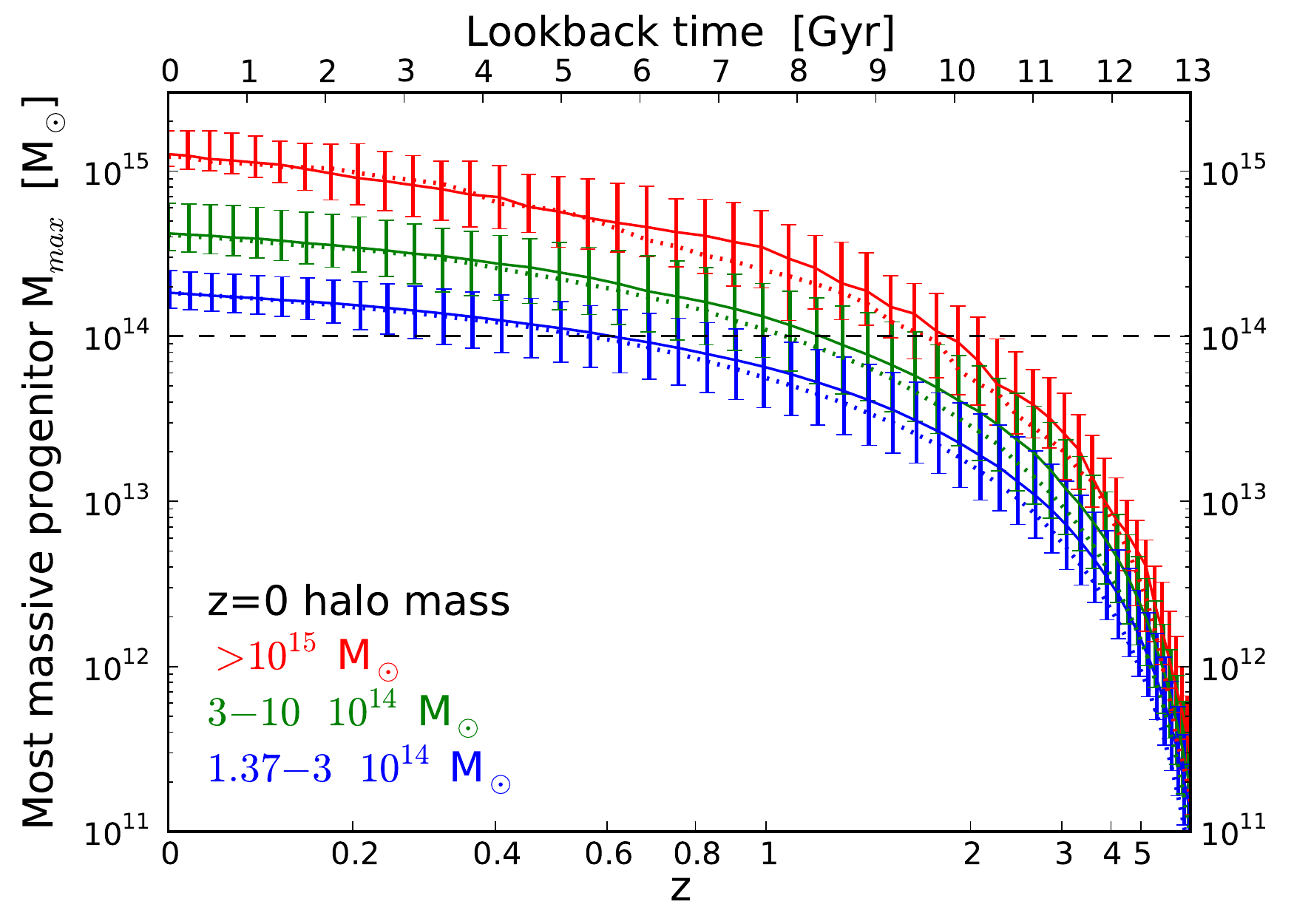}
\includegraphics[width=0.45\textwidth]{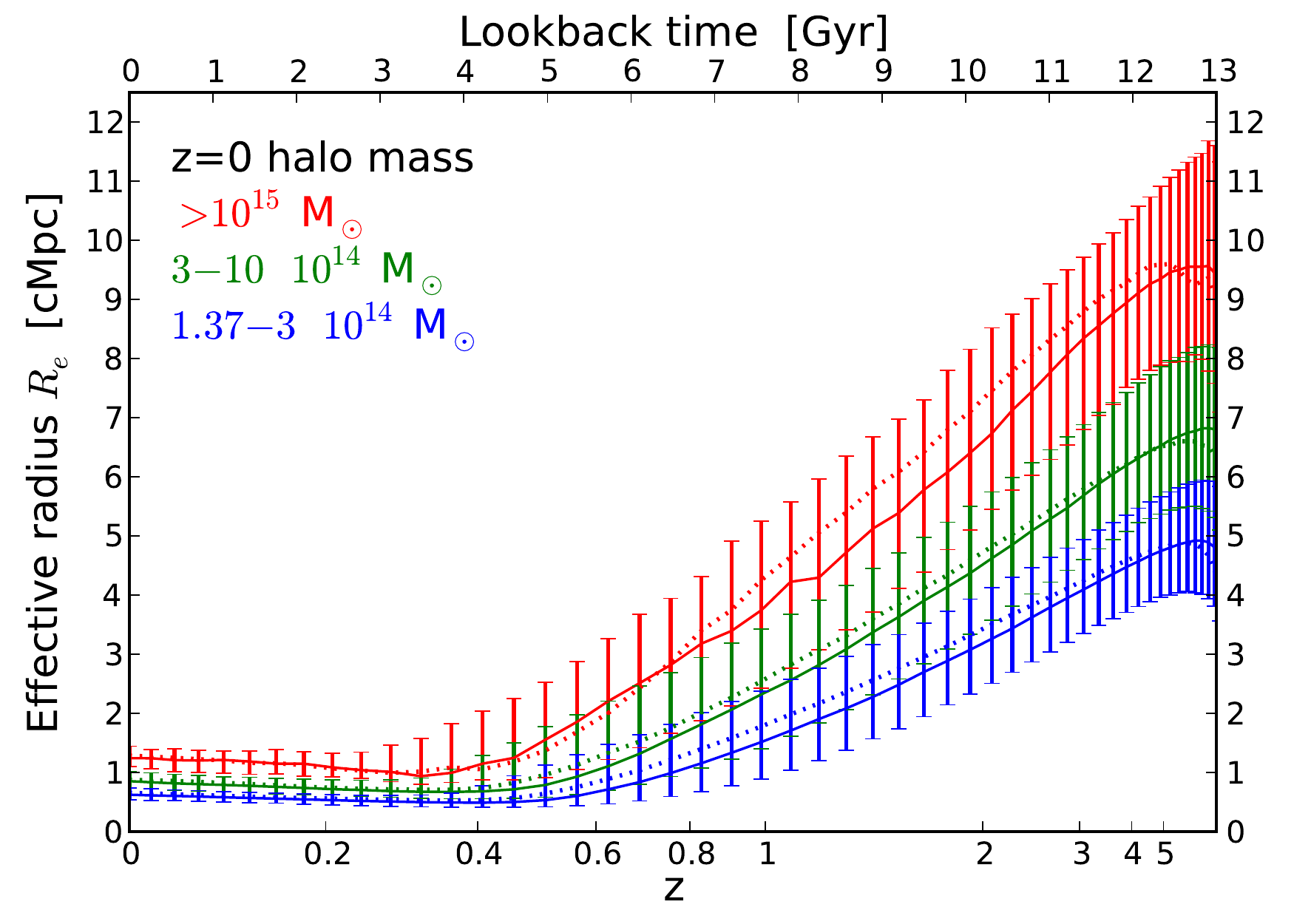}
\end{center}
\caption{\label{fig:pcsizes}Mass evolution for the most massive halo in a protocluster ({\it left panel}) and the size of a protocluster region ({\it right panel}) with redshift. The three colored lines indicate the median and $1\sigma$ scatter for Fornax- ({\it blue lines}), Virgo- ({\it green lines}), and Coma-type galaxy (proto-)clusters ({\it red lines}). The dotted lines indicate the results for the WMAP7 cosmology, instead of the WMAP1 cosmology. Figure reproduced from Fig. 2 in \citet{chiang13}.}
\end{figure}

\citet{chiang13} investigated in detail the properties of the high redshift overdensities associated with $\sim$3000 clusters sub-divided in poor (i.e., ``Fornax''-type or $1-3\times10^{14}$ $M_\odot$), average (i.e., ``Virgo''-type or $3-10\times10^{14}$ $M_\odot$) and rich (i.e., ``Coma''-type or $>10^{15}$ $M_\odot$) clusters found in the Millennium Run simulation. The expected size and mass evolution of these clusters with redshift are shown in Fig. \ref{fig:pcsizes}. The predicted comoving sizes of protoclusters (right panel) range from several to several tens of Mpc, in agreement with the observations. More massive clusters originate from larger cosmic volumes, and will thus, on average, have larger (projected) sizes on the sky. The evolution in halo mass for the most massive halo found in each (proto)cluster as a function of redshift is shown in the left panel of Fig. \ref{fig:pcsizes}. This reveals an important result, namely that even the most massive present-day clusters first acquired a halo mass of $\sim10^{14}$ $M_\odot$ only at $z\approx2$. \citet{wu13} showed that such clusters have an average half-mass redshift $z_{1/2}\sim0.6$ and experience their last major merger at $z_\mathrm{lmm}\sim1.6$. 

\begin{figure}[t]
\begin{center}
\includegraphics[height=0.4\textheight]{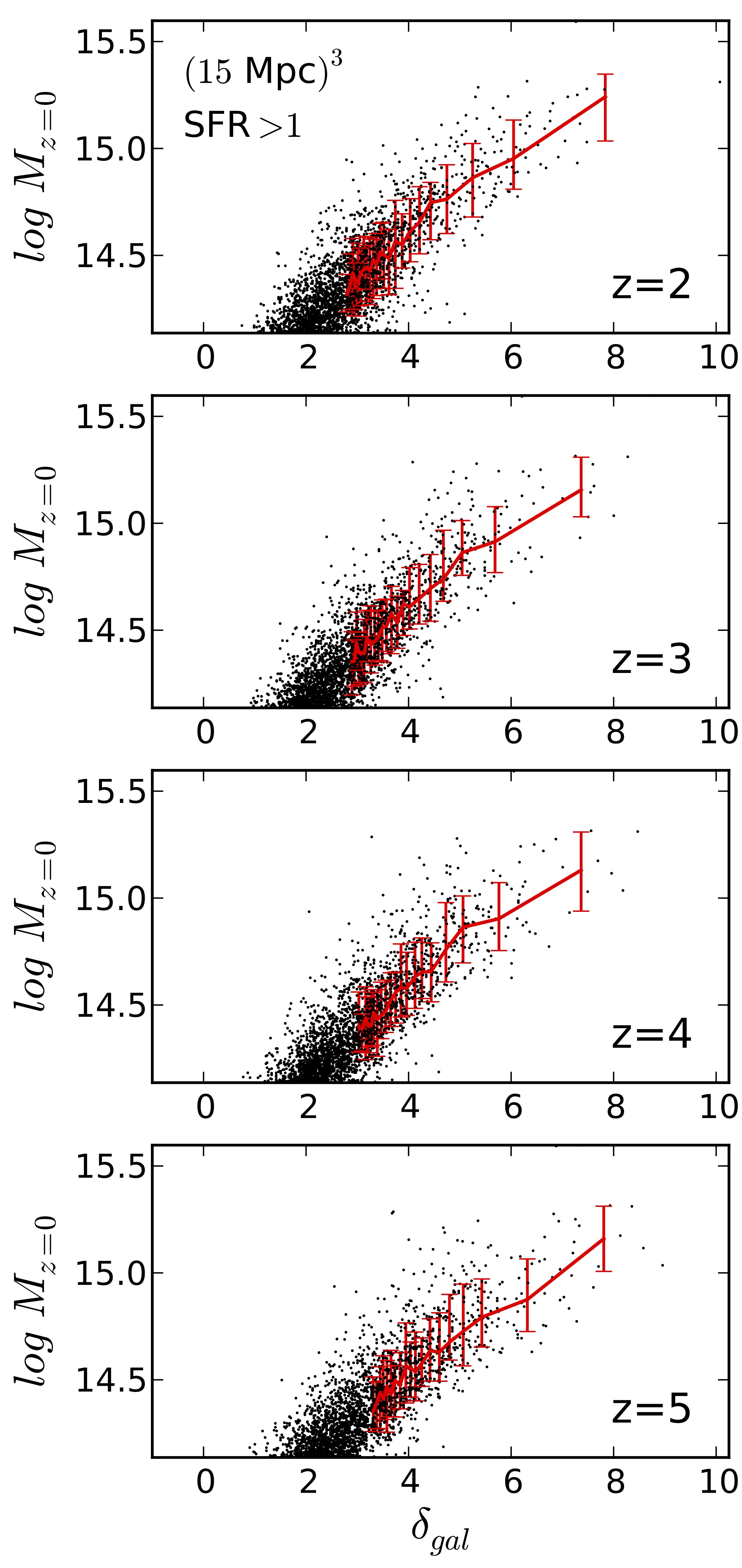}~~~~~
\includegraphics[height=0.4\textheight,width=6cm]{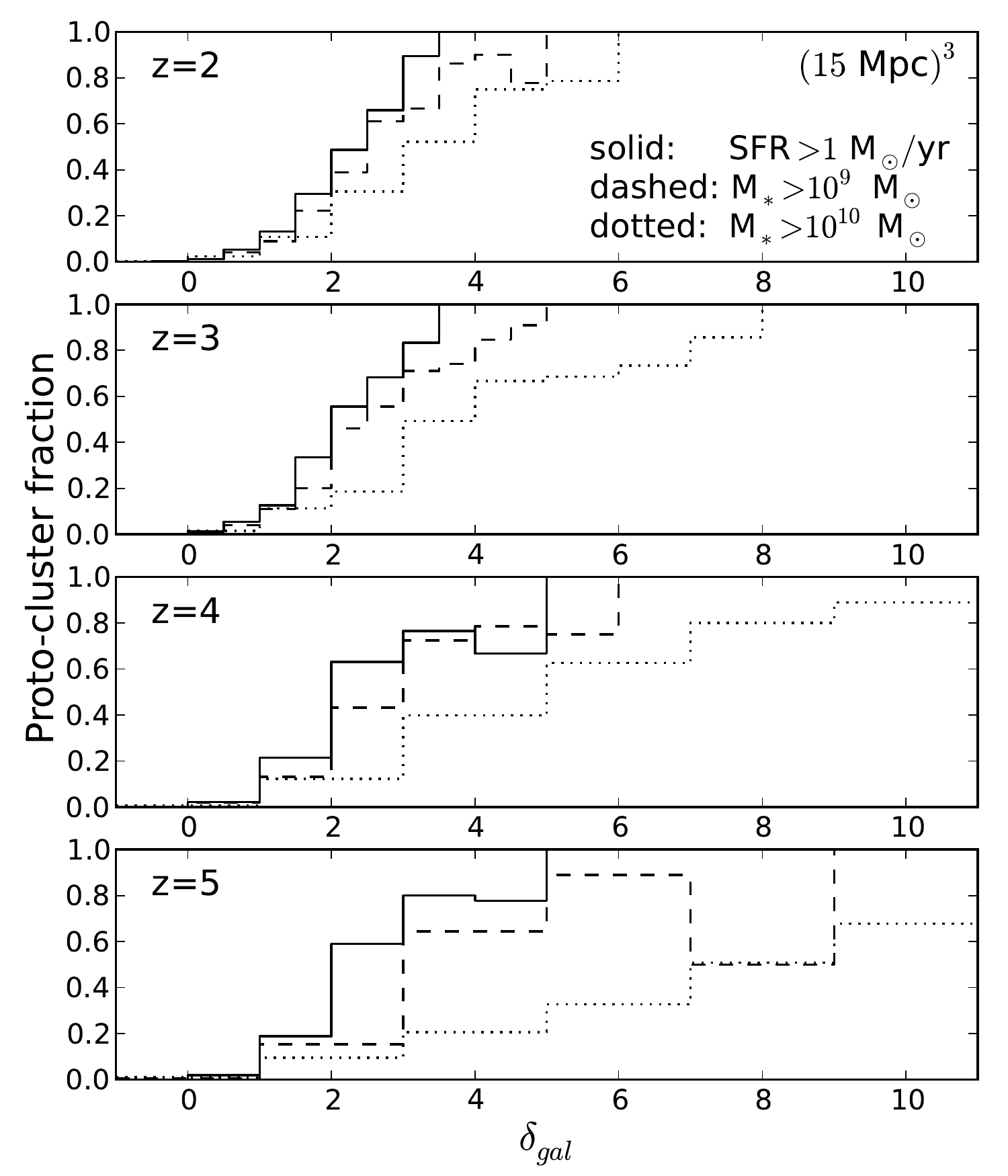}
\end{center}
\caption{\label{fig:chang13_dgal}{\it Left panel} Points indicate the relation between present-day cluster mass ($M_{z=0}$) and overdensity ($\delta_\mathrm{gal}$) measured for star-forming galaxies having a SFR $>1$ $M_\odot$ year$^{-1}$ (from {\it top-to-bottom} $z=2$, $z=3$, $z=4$ and $z=5$). The median relation and $1\sigma$ scatter are indicated in {\it red}. {\it Right panel} The fraction of regions of a given overdensity that will collapse to form a galaxy cluster. The different lines indicate the predictions for overdensities of galaxies having SFRs $>1$ $M_\odot$ yr$^{-1}$ ({\it solid}), and stellar masses of $>10^9$ ({\it dashed}) and $>10^{10}$ $M_\odot$ ({\it dotted}). Figures reproduced from Figs. 8 and 10 in \citet{chiang13}.}
\end{figure}

These kind of simulations are extremely powerful for interpreting the observational data of high redshift overdensities. \citet{chiang13} showed that there is a reasonably strong correlation between overdensities of galaxies measured at high redshift and the mass of their parent halo at the present day, at least out to $z\sim5$ \citep[see also][]{suwa06}. This is shown in the left panel of Fig. \ref{fig:chang13_dgal}. Progenitors of $\gtrsim10^{15}$ $M_\odot$ clusters can be identified by overdensities of star-forming galaxies of $\delta_\mathrm{gal}\gtrsim6$ when measured on $\sim15$ Mpc scales. The dispersion in the $\delta_\mathrm{gal}-\mathrm{log}M_{z=0}$ relation, as suggested by the simulation, amounts to a few tenths of dex in $\mathrm{log}M_{z=0}$. The fact that $\delta_\mathrm{gal}(z)$ is a fairly accurate indicator of the final cluster mass even during this pre-collapse phase of clusters, implies that it is indeed possible to compute the present-day cluster mass based on protoclusters observed at high redshift. In agreement with the results based on the spherical collapse model, present-day masses estimated based on the much more precise calibration between the observed properties of protoclusters and $z=0$ mass in numerical simulations are $\mathrm{log}[M/M]_\odot\simeq14-15.5$ \citep[e.g.,][]{governato98,suwa06,chiang15,lemaux14,cucciati14,hatch16a}. 

\citet{muldrew15} demonstrated that the evolutionary state of a protocluster could be determined better by computing the mass ratio of the most massive and second most massive protocluster halo. In the future, it may be possible to detect and weigh individual cores in a given protocluster based on galaxy velocity dispersions or deep X-ray observations. Alternatively, and perhaps easier to determine based on current data, one could measure the concentration of a protocluster to assess its state of collapse, analogous to the concentration parameter used in galaxy morphology studies. Two other very useful parameters that can be derived from these kind of simulations are the purity and completeness of protocluster samples. The purity is a measure of false positives of protoclusters. An example is shown in the right panel of Fig. \ref{fig:chang13_dgal}. As the overdensity increases, the fraction of regions with that overdensity that corresponds to protoclusters increases as well. At $z\sim3$, for example, it is near certain that an overdensity of $\delta_\mathrm{gal}\gtrsim8$ will be a protocluster, while an overdensity of $\sim1$ will in most cases not correspond to the progenitor of a galaxy cluster \citep{chiang13}. Another important quantity is the completeness. This is a measure of the fraction of protoclusters that are missed by a given selection. A single object found at some $z$ is a good (candidate) protocluster if the simulations indicate a high purity for the given selection function. A representative protocluster survey should aim to have both a high purity and a high completeness \citep[e.g.,][]{chiang14,toshikawa16}. 

\subsubsection{Other mass measurement methods}

Besides the spherical collapse and simulation-based methods described above, mass estimates have been derived in a number of different, complementary ways for a limited but growing number of protoclusters, such as dynamical masses, stellar mass-based masses and X-ray masses. \citet{lemaux14} found a dynamical mass of $\sim3\times10^{14}$ $M_\odot$ at $z=3.3$. Taking the average growth rate of such a halo between $z\sim3$ and $z=0$ then predicts a mass of $\sim9\times10^{15}$ $M_\odot$ for the $z=0$ cluster. Not only would this cluster be among the most massive clusters known today, it would also qualify as the highest redshift {\it cluster} already at $z\sim3.3$. As discussed by \citet{lemaux14} it is unlikely that the system is virialized, thereby invalidating these dynamical mass estimates. However, these dynamical mass estimates provide an upper limit for the actual masses, given that the galaxies most likely populate multiple halos within the protocluster system rather than one virialized system \citep[see also][]{dey16}. 
\citet{venemans07} found that the velocity dispersion of LAEs in radio galaxy protoclusters increases with decreasing redshift roughly as expected based on numerical simulations \citep[see also][]{contini15}. \citet{wang16} detected the core of protocluster CL J1001 at $z=2.5$ in the X-rays. Assuming that the X-rays are due to a thermalized ICM, they estimate a mass of $M_{200c}=10^{13.7\pm0.2}$ $M_\odot$, which is corroborated by its spectroscopic velocity dispersion of $510\pm120$ km s$^{-1}$. \citet{galametz13} detected a (proto)cluster with a velocity dispersion of $\sim400-600$ km s$^{-1}$ associated with the radio galaxy MRC 0156--252 at $z=2.02$ for which \citet{overzier05} previously detected extended soft X-ray emission with a flux $\sim2\times10^{13}$ erg s$^{-1}$ cm$^{-2}$ of uncertain origin. However, we must be cautious of X-ray detections of (proto)clusters at these redshifts. For example, radio galaxies at high redshift are often associated with extended X-ray emission that can be confused with an ICM \citep{carilli02,overzier05}. Furthermore, there is evidence that protoclusters have an elevated AGN fraction (Sect. \ref{sec:environmentaleffects}), which may bias the mass estimates based on low resolution X-ray data \citep[e.g.,][]{pierre12}. Another constraint on the cluster mass can be obtained by comparing the combined stellar mass of the galaxies in the protocluster with that of virialized clusters \citep[e.g.,][]{overzier09b,lemaux14}. Finally, as we have shown in Sect. \ref{sec:igm}, present-day cluster masses may also be estimated on the basis of \lya\ tomographic maps that have been calibrated against accurate IGM simulations.  

\section{Properties of protoclusters}
\label{sec:properties}

\subsection{Protoclusters as probes of structure formation}
\label{sec:evolution}

The evolution of galaxies in protoclusters may have diverged from that of galaxies in the field already at relatively early times. Protoclusters lie at the intersection of dense, gas-rich filaments in the cosmic web (Fig. \ref{fig:boylankolchin09}). It is well known that galaxies of higher stellar mass formed earlier than galaxies of lower stellar mass, and galaxies in high-density regions tend to form earlier than galaxies of the same mass in less dense regions \citep{thomas05}. Because the formation redshifts of cluster galaxies are high, we expect that (proto-)clusters should contain galaxies that are, on average, older and more massive compared to those in field environments at the same redshift. We may also expect differences between galaxies in and outside protoclusters, depending on the strength of  environmental effects occurring preferentially in overdense regions. Galaxies in present-day clusters experienced more major mergers compared to galaxies of a similar mass outside clusters \citep{gottlober01,fakhouri09}. Because merging activity stimulates both AGN and star formation activity and also largely determines the morphological and kinematical properties of galaxies, some environment-driven differences may thus be expected in protocluster galaxies even at redshifts before the more classical environmental effects such as ram-pressure stripping or dynamical friction become important. 
\begin{figure}[t]
\begin{center}
\includegraphics[width=\textwidth]{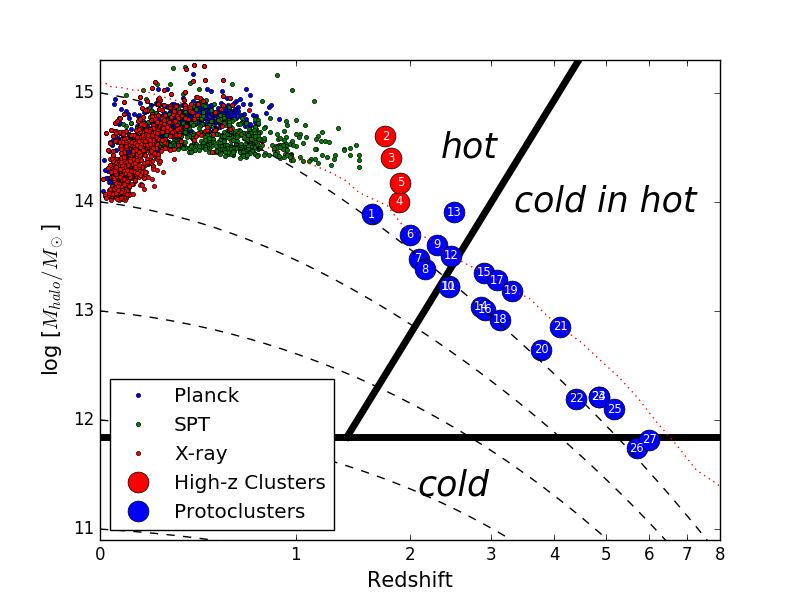}
\end{center}
\caption{\label{fig:mydekel}
The accretion history of (proto)clusters. Between $z\approx2$ and $z\approx3$ the core halos of protoclusters are expected to transition from a regime where the inflow of cold gas is still possible in massive halos (``cold in hot'') to a regime where all gas accreted onto massive halos shock-heats to the virial temperature (``hot''). The estimated masses of clusters and protoclusters above $z=1.5$ are indicated by large numbered {\it symbols}. The number in each symbol corresponds to the references used for the data (1: \citet[][]{papovich10,pierre12}; 2: \citet[][]{stanford12,brodwin12}; 3: \citet[][]{newman14}; 4: \citet[][]{zeimann12}; 5: \citet[][]{mantz14}; 6: \citet[][]{gobat11,gobat13,strazzullo13,valentino15}; 7: \citet[][]{spitler12,yuan14}; 8: \citet[][]{venemans07,kurk04a}; 9: \citet[][]{steidel05}; 10: \citet[][]{chiang15}; 11: \citet[][]{diener15}; 12: \citet[][]{casey15}; 13: \citet[][]{chiang14,casey15,wang16}; 14: \citet[][]{venemans07}; 15: \citet[][]{cucciati14}; 16: \citet[][]{venemans07}; 17: \citet[][]{steidel98}; 18: \citet[][]{venemans03}; 19: \citet[][]{lemaux14}; 20: \citet[][]{kslee14,dey16}; 21: \citet[][]{venemans02}; 22: \citet[][]{kuiper11b}; 23: \citet[][]{shimasaku03}; 24: \citet[][]{matsudarichard10}; 25: \citet[][]{venemans07}; 26: \citet[][]{ouchi05}; 27: \citet[][]{toshikawa14}). Cluster data below $z=1.5$ were taken from the compilation of clusters in X-ray and SZ surveys of \citet{bleem15}. The red dotted line traces the mass of the most massive halo of a Coma-type cluster \citep{chiang13}. Dashed lines trace the average halo mass growth based on the fitting functions of \citet{behroozi13}. Thick black lines indicate the different gas cooling regimes as predicted by \citet{dekel06}. 
}
\end{figure}

The large-scale inflow patterns of the gas flowing into protocluster regions is also expected to be very different from that around galaxies in more average regions. The early onset and enhancement in star-formation and AGN activity in protoclusters will furthermore affect the outflow patterns of warm gas and the redistribution of metals on large scales mixing with fresh inflowing gas, while the growth of the massive cluster-sized halo will establish a region of hot gas extending out to the virial radius of the halo. The centers of protoclusters should thus be surrounded by a very complex multi-phase medium not typically encountered elsewhere. \citet{dekel06} showed that below a critical shock-heating mass of $M_\mathrm{shock}\sim10^{12}$ $M_\odot$, galaxies are built by streams of cold gas (in addition to merging) that are not interrupted by virial shock-heating. For more massive halos above the shock-heating mass the situation is more complicated: at low redshift ($z\lesssim2$) gas accreted from the IGM will immediately shock-heat to the virial temperature of the halo and will generally only be available for star formation after relatively long cooling times. At high redshift, however, cold gas is still presumed to be able to penetrate the hot gas environments in massive halos through so-called cold flows that were established prior to the formation of the shock-heated gas in these halos \citep[see also][and references therein]{keres05,keres09,ocvirk08}. This general scenario has been shown to be able to reproduce many of the most essential features of galaxy evolution, such as the relatively high SFRs observed in massive galaxies at high redshift, the origin of the present-day galaxy bimodality, and downsizing \citep[e.g.,][]{dekel06,cattaneo07,cattaneo08}, although many details are still unknown and in need of observational evidence. 

We propose that protoclusters are uniquely suited for testing several crucial and currently unconstrained aspects of this scenario. This is demonstrated in Fig. \ref{fig:mydekel}. Here we have made a first attempt to place the existing data on high redshift clusters and protoclusters from Fig. \ref{fig:pcredshifts} on an evolutionary sequence in terms of halo mass and redshift. Accurate mass estimates exist for several of the clusters in the $1.5\lesssim z\lesssim2$ redshift range based on X-rays, SZ effect or velocity dispersions (large red symbols in Fig. \ref{fig:mydekel}). In the case of the $z=2.5$ protocluster studied by \citet{wang16} an estimate for the mass of its massive core halo also exists based on X-rays and velocity dispersions (assuming that both trace the virialized component). For most other protoclusters it is much harder to place them on the diagram as no such mass estimates exist. However, we can estimate the mass of the most massive halos present in each protocluster by taking their estimated $z=0$ cluster mass and evolving it back in time (large blue points). To perform this extrapolation we used the mean relations between the mass of the most massive halo present in a (proto)cluster at each redshift for a given $z=0$ cluster mass based on the simulations of \citet{chiang13}. Although this extrapolation is likely to be highly uncertain for each individual protocluster given the large uncertainties in their $z=0$ mass and relatively small number statistics, here we are mostly interested in showing the general trends based on the ensemble as a whole. We have also indicated the approximate gas cooling regimes expected for halos of a given mass and redshift [the regimes are marked ``hot", ``cold" and ``cold in hot" and are separated by the thick lines; from \citet{dekel06}]. 

It can be seen that protoclusters occupy a very interesting region in this diagram. The shock-heating mass $M_\mathrm{shock}\sim10^{12}$ $M_\odot$ is believed to be relatively independent of redshift, as it relates to the halo mass above which an extended stable shock can expand to the virial radius. However, the typical mass of the halos transitioning from the cold to the hot regime have an important redshift dependence: halos of $\sim10^{12}$ $M_\odot$ make the transition only at $z\sim1$, while for halos of $\sim10^{13}$ $M_\odot$ ($\sim10^{14}$ $M_\odot$) the transition is expected to occur as early as $z\sim2$ ($z\sim3$). The central regions of protoclusters are believed to contain the first massive halos that make the transition from the cold flows regime into the hot regime. The average (proto)cluster shown in Fig. \ref{fig:mydekel} should make this transition at $z\sim2.5$ when their core halos had masses of a few times $10^{13}$ $M_\odot$. There are many high redshift clusters and protoclusters on either side of the diagonal line in Fig. \ref{fig:mydekel} separating the two regimes. We may thus expect important changes to occur in these environments as these systems evolve from the protocluster to the cluster regime (see Sect. \ref{sec:finalremarks} for more discussion).  

\subsection{Evidence for a red sequence in protoclusters?}
\label{sec:redsequence}

The formation of the galaxy red sequence is an extremely important and unsolved problem in galaxy evolution not exclusive to (proto)clusters. It is well established that the fraction of massive ($\gtrsim10^{11}$ $M_\odot$) galaxies that is quiescent is relatively constant at $0<z<1$, while the stellar mass density of quiescent galaxies of lower mass increases with decreasing redshift both in the field \citep{ilbert13} and in galaxy clusters \citep{gilbank08}. This quenching could be related to feedback from AGN or to the build-up of hot gas reservoirs around massive galaxies in field and clusters alike. However, it has been demonstrated that there also needs to be an environmental component to the quenching process, particularly in the dense environments of galaxies in groups and clusters. The local trend of higher quiescent fractions in more denser environments or more massive halos continues toward higher redshifts, although it is considerably weaker at the highest redshifts considered \citep[$z\sim1-2$;][]{quadri12,scoville13,lee15}. Massive clusters at $z\sim1$ generally have strong red sequences \citep[e.g.,][]{postman05,blakeslee06,vanderwel07,patel09,mei09,muzzin12,foltz15}. Although the ages of the quiescent galaxies in high redshift clusters appear to be comparable to those in the field \citep[e.g.,][]{rettura10,gobat08,newman14}, the star formation timescales for galaxies inside clusters may be more contracted \citep{rettura11}. Massive red galaxies and high quiescent fractions of massive cluster members continue to be found at $1.5<z<2$ \citep[e.g.,][]{tanaka10a,papovich10,rudnick12,fassbender14,vanderburg13,strazzullo13,andreon14,newman14,koyama14,cooke16,hatch16b}, although these systems also frequently host star-forming or starbursting galaxies not seen at the lower redshifts \citep[e.g.,][]{hilton10,tran10,fassbender11,gobat13,brodwin13,strazzullo13,lee15,mei15,wang16} \citep[but see][]{cooke16,andreon14,newman14}. 

The evolution of the cluster red sequence is the net result of several competing or complementary processes responsible for either driving or shutting down star formation in galaxies. The dominant mechanisms are expected to differ across redshift, stellar mass, halo mass and environment, and it is still a challenge for models to reproduce the quiescent cluster galaxy population observed at high redshift \citep[e.g.,][]{sommariva14}. Galaxy (proto-)clusters offer a unique chance to investigate the red sequence formation because they allow us to study dense regions out to higher redshifts, i.e., closer to the epoch of formation and quenching of the massive cluster galaxies. 
What is the evidence for red sequence galaxies in the sample of protoclusters available to date? One of the first objects that allowed us to address this question is the protocluster associated with the Spiderweb galaxy, a radio galaxy  at $z=2.2$. \citet{kurk04b} identified an excess of photometrically selected extremely red objects (EROs) coinciding with the protocluster. They concluded that most of the EROs were likely to be dust-enshrouded starburst galaxies but marked 5 galaxies as potential quiescent red sequence objects. \citet{zirm08} performed the first deep near-IR study of this field, using HST/NICMOS to search for objects having relatively red $J-H$ colors (approximately rest-frame $U-B$) indicative of a forming red sequence. Although still working with photometrically selected (i.e., not spectroscopically confirmed) samples, they found a factor of 6 overdensity of relatively red objects over a small area surrounding the protocluster compared to similar observations performed in the HUDF. They showed that these galaxies were bluer than galaxies on the red sequence in both the Coma cluster and in RDCS 1252.9--2927, a cluster at $z=1.24$ used for reference. However, the objects were redder than expected based on passively de-evolving the RDCS 1252.9--2927 red sequence to $z=2.2$ [this assumed a formation redshift of $z\sim3$ based on the analysis of \citet{blakeslee03}]. Although this could indicate that the Spiderweb protocluster contains a passive population older than that expected for an RDCS 1252.9--2927-like cluster, it cannot be ruled out at the moment that the red excess is due to dusty star-forming galaxies. \citet{zirm08} also showed that their brightest and reddest objects have high Sersic indices ($n=2.7-5$) indicative of early-type morphologies, which supports the conclusion that at least some of these objects are genuine quiescent galaxies rather than dusty galaxies. \citet{kodama07} explored the red populations toward 4 regions at $2\lesssim z\lesssim3$, previously identified as protoclusters based on excesses of LAEs and HAEs near radio galaxies. They found that the bright, red end of the CMRs are more populated with respect to comparison field data. The excesses are particularly strong for the targets at the low redshift end ($z\lesssim2.5$) of their small sample, indicating that a red sequence of massive, passively evolving galaxies was first established between $z\sim2$ and $z\sim3$ (with $z_\mathrm{f}\simeq3.5-5$). Similar results based on near-IR color cuts in the fields of other radio galaxies at these redshifts were obtained also by other authors \citep[e.g.,][]{kajisawa06,hatch11a}. 

Although the excesses of red galaxies observed in these structures are promising, we must be cautious not to overinterpret these results as they are based on photometric selections only. \citet{doherty10} obtained a large number of optical and near-IR spectra of red galaxies in two protoclusters that had previously been shown to have rather prominent excesses of red galaxies, including the Spiderweb \citep{kurk04b,kodama07,zirm08}. Out of 90 objects targeted, membership of the $z=2.2$ cluster could be confirmed for just two red objects. Of these, one object was found to be a star-forming galaxy reddened by dust (based primarily on a 24 $\upmu$m detection), while the other could be a genuine quiescent galaxy (based on a low SFR based on \ha\ and a high stellar mass based on SED fitting). Both objects were found to have stellar masses of a few times $10^{11}$ $M_\odot$, and are thus suitable candidates for becoming massive red sequence cluster galaxies \citep[unless they merge with the BCG, see][]{doherty10}. By combining photometric data with low S/N spectroscopic data \citep[e.g.,][]{kriek06}, \citet{tanaka13} were able to further constrain the redshifts and physical properties of the red population. Assuming protocluster membership, they showed that there exists a small population of massive, quiescent galaxies (sSFR $<$ 10$^{-11}$ year$^{-1}$) consistent with a formation redshift of $z_\mathrm{f}\simeq3-4$ and a relatively short star formation timescale ($\tau\lesssim0.5$ Gyr). 

It may be clear from the above that the study of red sequence formation in protoclusters is severely limited by the lack of spectroscopic redshifts available for the candidate red cluster members. In part this is due to protocluster surveys and selection techniques relying often on star-forming galaxy populations rather than on quiescent galaxies to identify the protoclusters, and in part due to the difficulty in obtaining large numbers of spectroscopic redshifts of relatively quiescent galaxies at high redshift. One exception is the study of \citet{lemaux14} that found a highly overdense region ($\delta_\mathrm{gal}\approx11$) in a large spectroscopic survey (see Sect. \ref{sec:why}). VUDS is a relatively wide-field ($\approx3$ deg$^2$) survey that targets galaxies having a high likelihood of being at high redshift ($2\lesssim z\lesssim6$) based on their photometric redshifts, rather than using a magnitude cut. At the redshift of the protocluster Cl J0227--0421, the VUDS survey is sensitive to galaxies down to a stellar mass limit of $\sim10^9$ $M_\odot$, allowing an investigation of the colors of spectroscopically confirmed protocluster and field galaxies as  a function of stellar mass. The most notable finding is a significant excess of relatively red galaxies having high stellar masses of $\sim10^{11}$ $M_\odot$ in the protocluster compared to the field. Although the absolute number of massive, red galaxies is small (three, excluding the proto-BCG which is equally massive but has blue colors), they represent a factor $\sim$25 overdensity compared to the field. The excess is still a factor of $\sim$3 when taking into account that the protocluster itself is overdense compared to the field. \citet{lemaux14} refer to this population as ``proto-red sequence galaxies'' as they have masses consistent with those of red sequence galaxies in low redshift clusters, and colors and stellar populations that indicate no recent major star formation since $\sim$300 Myr, even though they are not strictly ``passive''. This implies formation redshifts of these stellar populations of $z\sim4$, i.e., similar to that derived for massive cluster galaxies at low redshift. The excess of massive galaxies as found in Cl J0227--0421 confirms the trend seen in many protoclusters in the redshift range $z\sim2-3$, but the additional excess of relatively red, possible (proto-)red sequence galaxies, has so far been unambiguously demonstrated in only very few systems \citep[e.g.,][]{lemaux14,wang16} \citep[but see][for counter evidence also based on a large number of spectroscopic redshifts in the zCOSMOS and VUDS surveys]{diener13,cucciati14}.  

\citet{contini15} investigated the growth of the red sequence in protoclusters having present-day masses of $\sim10^{15}$ $M_\odot$ in the context of semi-analytic models. These authors predict that at $z\sim3$, $\sim$90 \% of the progenitors of cluster red sequence galaxies are still actively star-forming. The passive red sequence forms primarily through the quenching of satellite galaxies as they deplete their cold gas reservoir after being accreted onto the cluster. Satellite galaxies accreted at higher redshifts ($z\simeq2-3$) have higher SFRs and quench faster compared to satellites that are accreted at lower redshifts ($z\simeq1-2$). This is because galaxies at higher redshifts have higher gas fractions and lower mass halos where outflows are more effective at removing cold gas. For central galaxies, the quenching occurs much later ($z<1$, with $\sim$20 \% quenched by $z=0.5$), and is primarily caused by AGN feedback that offsets the cooling of the hot gas. The models of \citet{contini15} predict that the quiescent fraction in (proto)clusters is no more than 20 \% at $z\sim3$, about 30 \% at $z\sim2$ and about 60 \% at $z\sim1$. These numbers appear to be consistent with the data for protoclusters as well as most $z<2$ clusters, except perhaps for JKCS 041 at $z=1.80$ which has an extraordinary quiescent fraction of 80\% \citep{newman14}. This may indicate that cosmic variance or assembly bias are important factors when studying (proto)clusters at these redshifts, highlighting the importance of increasing the sample size. \citet{newman14} found no differences in the ages of the stellar populations of the red sequence galaxies compared to a sample of co-eval galaxies in the field. This suggests that the epoch of quenching of star-formation in massive galaxies is similar for clusters and the field, although the process may be more efficient in clusters given the large number of quenched galaxies found, at least in JKCS 041 \citep{newman14}. 

\subsection{Brightest cluster galaxies} 
\label{sec:bcg}

Brightest cluster galaxies are the most massive galaxies at the centers of galaxy clusters. As shown in \citet{lin04}, they are typically located close to the cluster center as defined by the peak X-ray surface brightness. They are more massive in more massive clusters, and contain a smaller fraction of the total light with increasing cluster mass. BCGs accumulate their mass primarily through mergers, and the mass fraction of stars contained in the ICM, believed to be a by-product of such mergers, increases with increasing cluster mass \citep{lin04}. The largest type of BCGs, the `cD' galaxies, often contain secondary or multiple nuclei indicative of their cannibalistic past \citep[see][]{tonry87}. BCGs have several important structural and chemical properties distinct from non-BCGs of the same stellar mass, and more frequently host radio-loud AGN \citep{vonderlinden07,best07}. Their formation history is not only governed by the evolution of massive galaxies in general, but likely also depends on the complexities of cluster formation and significant environmental effects. By comparing the evolution of BCGs as observed in matched cluster samples at different epochs with the evolution predicted by simulations, we can test both the large-scale cluster assembly process and the detailed assembly process of the BCG in the cluster centers \citep[e.g.,][]{delucia07,collins09,lin13,inagaki15,zhao15}. 

\begin{figure}[t]
\begin{center}
\includegraphics[width=0.45\textwidth]{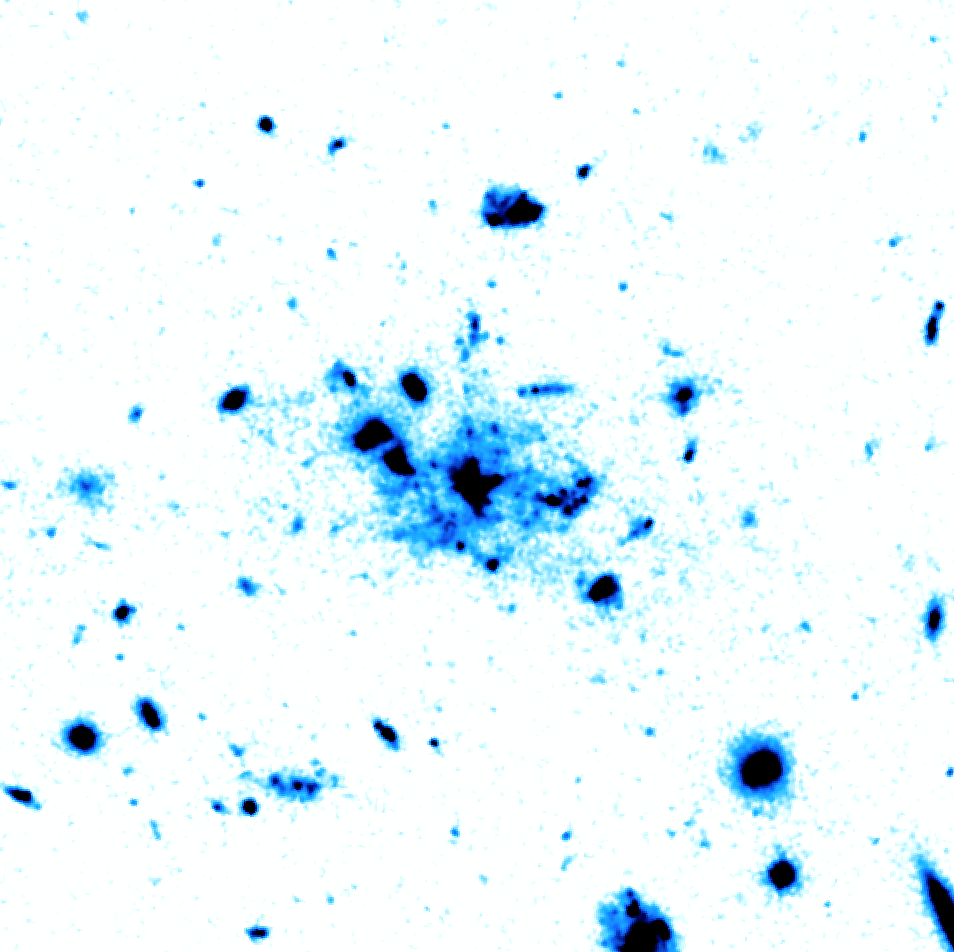}~~~
\includegraphics[width=0.45\textwidth]{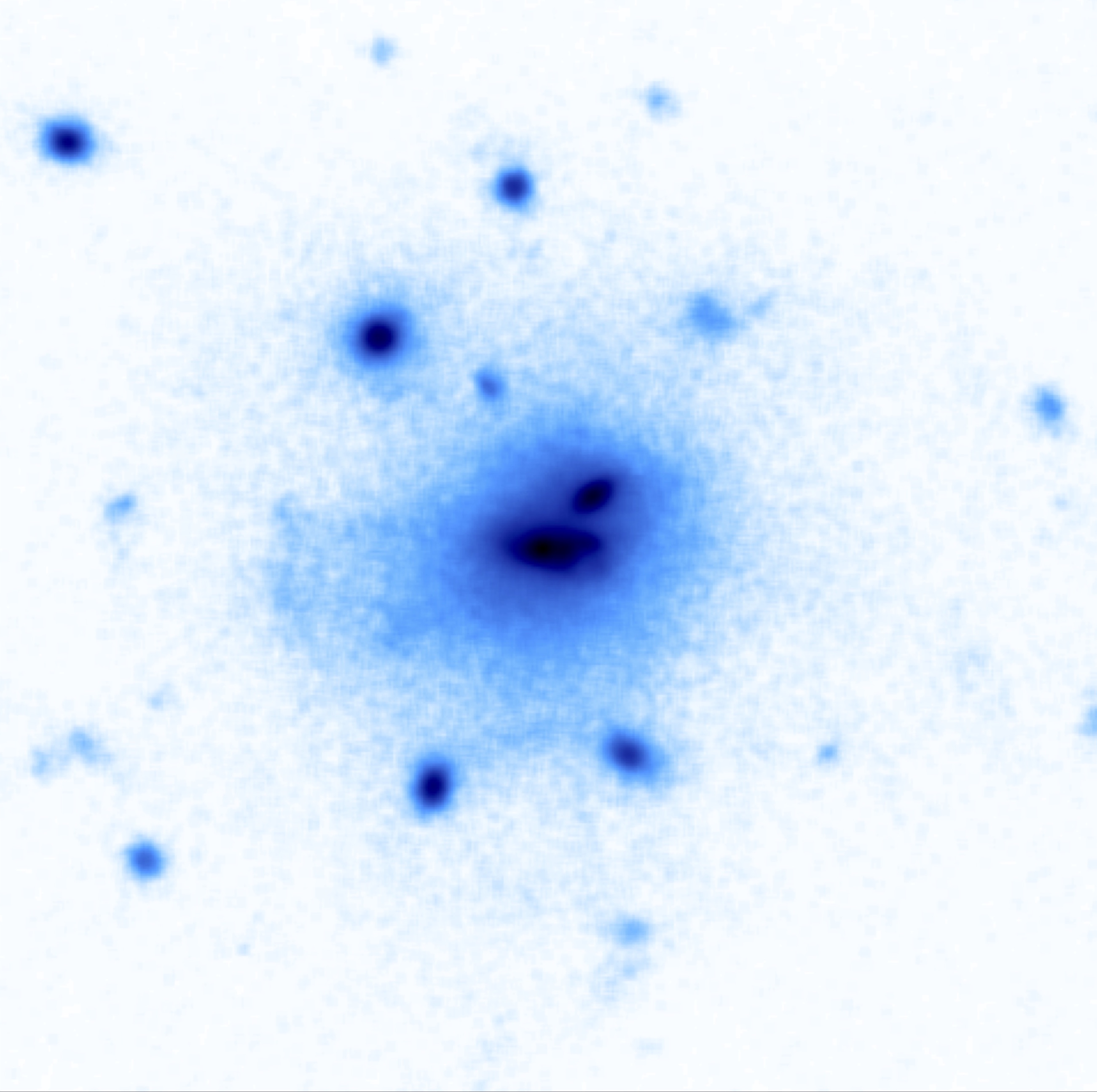}
\end{center}
\caption{\label{fig:bcg_sim}Observations and simulations of proto-BCGs. Left panel: rest-frame UV image of the Spiderweb Galaxy at the center of a $z\approx2.2$ protocluster \citep[][]{miley06}. Right panel: simulated image of the stellar mass density around a proto-BCG in a forming galaxy cluster at $z=2.1$ \citep{saro09}. The excesses of small and large satellites, the centrally dominant massive galaxy and the diffuse extended nebula of stars are in qualitative agreement with the properties of the Spiderweb Galaxy shown {\it left}. The images are shown at the same scale, and measure 150 $h^{-1}$ physical kpc in diameter. The image shown in the right panel was taken from Fig. 1 in \citet{saro09}.}
\end{figure}

Protoclusters offer a unique opportunity to observe the initial stages of the development of BCGs, which can provide some much needed constraints for models and numerical simulations that are relatively untested at these high redshifts. It is expected that BCG formation will be accompanied by excessive merging, star formation and AGN activity in the dense protocluster environment, especially at high redshift. The general class of high redshift radio galaxies have many properties indicative of progenitors of local BCGs \citep[e.g.,][]{pentericci97,best98,zirm05,miley08,hatch09,overzier09b,collet15}. Perhaps one of the most spectacular examples of such a proto-typical BCG found to date is the `Spiderweb galaxy', a radio galaxy at the center of a $z=2.2$ protocluster. 
This system has many intriguing features that both quantitatively and qualitatively suggest that it is a forming BCG (see Figs. \ref{fig:bcg_sim}, \ref{fig:labs}). Its main component in the rest-frame optical has a classical De Vaucouleurs profile, while its larger-scale morphology in the rest UV and optical is extremely clumpy and contains numerous faint satellites and diffuse emission \citep{pentericci97,pentericci98,miley06}. It has an exceptionally large stellar mass \citep[$\sim10^{12}$ $M_\odot$;][]{pentericci97,hatch09}, a high molecular mass \citep[$M_{H_2}\sim10^{11}$ $M_\odot$;][]{emonts13}, a high SFR \citep[$>1000$ $M_\odot$ year$^{-1}$;][]{stevens03,hatch08,seymour12}, and an SMBH of $\sim10^{10}$ $M_\odot$ \citep{nesvadba11a}. The radio AGN shows evidence for strong interaction with the surrounding medium \citep{pentericci97}, capable of removing a large fraction of the gas and quenching star formation \citep{nesvadba06}. On the scale of present-day cD galaxy envelopes, the system shows a luminous emission line nebula \citep{pentericci97,kurk00}, extended star formation resembling a young version of the ICL \citep[][]{pentericci98,hatch08}, a dense (magnetized) gaseous medium with high Faraday rotation measure \citep{pentericci97}, and extended soft X-ray emission due to shocked gas or a faint thermal ICM \citep{carilli02}. \citet{saro09} performed a set of hydrodynamical simulations that followed the growth of the BCG in poor and rich clusters (Fig. \ref{fig:bcg_sim}). By comparing the SFR, mass and morphology of the BCGs and the luminosity functions and velocity dispersions of the protoclusters, \citet{saro09} found that the properties of the Spiderweb protocluster span the range of properties between the two simulations. They also showed that these systems are already surrounded by a significant fraction of ICL resulting from the early merging activity of the proto-BCG. 

Although the Spiderweb galaxy appears to be relatively exceptional compared to other systems known at the present moment, several other protoclusters show evidence for BCGs in various stages of formation, such as massive, quiescent galaxies at the centers of dense groups in protoclusters \citep[e.g.,][]{spitler12,yuan14,belli14,hatch16b}, tight groups of central interacting galaxies \citep{gobat11,gobat13,strazzullo13}, massive AGN host galaxies \citep{lemaux14,hennawi15}, and dense groups of massive quiescent or (dusty) star-forming galaxies having dynamical or total stellar masses consistent with those expected for present-day BCG progenitors \citep[e.g.,][]{kubo16,ishigaki16}. However, it is also important to note that there are many examples of protoclusters in which no particular galaxy members are present, or in which they have not yet been identified \citep[e.g.,][]{cucciati14,steidel05,ouchi05,toshikawa14,dey16}. 

It is tempting to draw conclusions regarding the early evolution of BCGs by connecting, e.g., the Bootes protocluster at $z=3.78$ that contains no particular members \citep{kslee14,dey16}, the compact group of massive and starbursting galaxies in the core of the SSA 22 protocluster at $z=3.1$ \citep{kubo16}, the centrally dominant and highly active `Spiderweb Galaxy' at $z=2.2$, and the relatively quiescent systems seen in clusters at $z\lesssim2$. However, given the small sample sizes and the hugely varying selection techniques used, we must be very careful not to overinterpret the data at this early stage. It can be instructive to compare the observational evidence with simulations of BCG formation. \citet{delucia07} used a semi-analytic model to specifically study BCG formation \citep[see also][]{dubinski98,gao04a}. One of the key predictions of that study is a large discrepancy between the assembly history (defined as the mass growth of the main progenitor of the BCG with time) and the star formation history (SFH) of the BCG (defined as the sum of the SFHs of all the progenitor galaxies that merge to form the BCG). More specifically, they found that, on average, 50 and 90 \% of the stars in present-day BCGs formed at $z\gtrsim4$ and $z\gtrsim2$, respectively. In comparison, the redshifts at which 50 and 90 \% of the stellar mass in these BCGs were assembled into a single galaxy are much lower, at $z\sim0.5$ and $z\lesssim0.2$, respectively. Prior to $z\sim1$, the main BCG progenitor is typically difficult to identify, especially since it is not necessarily the most massive galaxy in the protocluster. This picture is, at least qualitatively, consistent with those protoclusters in which no obvious proto-BCGs have been identified, or those in which there appear to be several massive (central) objects. Although the relatively late assembly of BCGs as suggested by these simulations appears to be in disagreement with protoclusters that already show evidence for a significant central object \citep[e.g.,][]{pentericci97,overzier09b,lemaux14,hennawi15}, it is worth highlighting another important feature of the models: when studying the properties of the central galaxies in the most massive haloes selected at high redshift rather than at the present-day, \citet{delucia07} and \citet{angulo12} found that these are generally not the progenitors of the most massive haloes at $z=0$ (although they are still found in clusters). It is not unlikely that the biased tracer technique used to find protoclusters near radio galaxies and quasars preferentially selects systems in which the central galaxies are in a more advanced stage compared to those in the more typical progenitors of clusters of the same mass. This could be tested with larger samples of protoclusters and a better understanding of the selection effects.    

These massive (forming) galaxies in protoclusters offer new insights into the progenitors of the BCGs in local and high redshift clusters. Common characteristics include high stellar masses, high SFRs, significant AGN activity and feedback, complex systems in the process of merging, and extended reservoirs of ionized or molecular gas. As the sample of high redshift clusters and protoclusters expands in the coming decade, we will be able to track the growth of the BCGs in these systems and provide valuable input for models of the formation of this important class of galaxies. 

\subsection{Other aspects of protocluster galaxies}
\label{sec:environmentaleffects}

The protocluster data available today, although still limited to a few well-studied systems, are starting to offer us a unique chance to study galaxy formation in a regime that was, until recently, only accessible in theory and simulations. Here we will briefly review some observational results related to the morphologies, stellar populations, gas-phase abundances and AGN fraction of galaxies in protoclusters. 
\begin{figure}[t]
\begin{center}
\includegraphics[width=\textwidth]{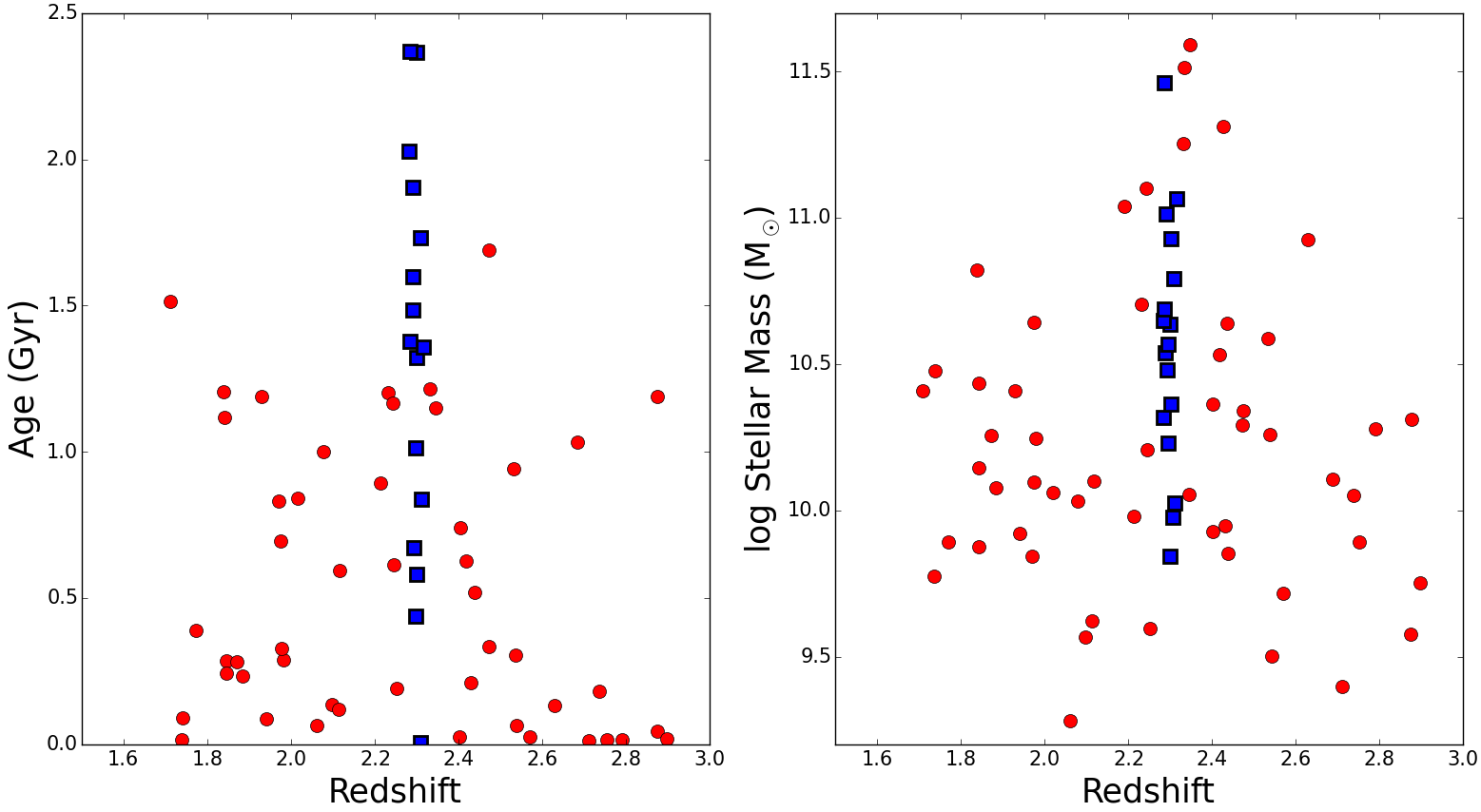}
\end{center}
\caption{\label{fig:steidel05}Stellar ages ({\it left panel}) and stellar masses ({\it right panel}) of the UV-selected galaxies in and outside the protocluster HS1700+643 at $z=2.30$ discovered by \citet{steidel05}. The protocluster contains galaxies that are both more massive and older than galaxies at neighboring redshifts. Data taken from Fig. 3 in \citet{steidel05}.}
\end{figure}

\subsubsection{Morphologies of protocluster galaxies} 

Protoclusters not only allow us to study massive galaxy formation in general (protoclusters contain large numbers of co-eval galaxies), they have also allowed some of the first quantitative studies of galaxy morphology in dense regions at $z\gtrsim2$. Some protoclusters were extensively imaged with HST, providing size and morphology measurements in the rest-UV for large numbers of LAEs and LBGs \citep[e.g.,][]{miley04,venemans05,overzier06,overzier08,peter07}. These studies showed the typical range of structures known for these kind of objects consisting of single, double, or multiple knotty structures with relatively small effective radii of, on average, $\sim1$ kpc. The sizes of the LAEs are generally smaller than those of the LBGs, which reflects the fact that they are also fainter and younger galaxies, but this is no different in protoclusters and in the field \citep{venemans05,peter07,overzier08}. \citet{miley06} observed a higher frequency of chain and tadpole morphology star-forming galaxies close to the Spiderweb Galaxy, the presumed proto-BCG of a protocluster at $z=2.2$, compared to the field. \citet{hine16} found an enhanced major merger rate in the $z=3.1$ SSA 22 protocluster based on visual classification and pair counts in the rest-UV. Such an enhanced merger rate could lie at the basis of a number of other environmental effects seen in this region, such as an enhanced AGN fraction \citep{lehmer09} and an excess of both star-forming and quiescent massive galaxies \citep{kubo13,kubo15,casey16}. 

Some studies have also reported that early-type galaxies in the highest redshift (proto-)clusters appear larger than those in the field. In principle, some degree of accelerated structural evolution may be expected given the much higher pair counts and inferred merger rates in the overdense environments of (proto-)cluster galaxies \citep{lotz13}.
Consistent with such a higher merger rate, \citet{zirm12} find an indication that massive, quiescent galaxies in a $z=2.2$ protocluster are larger, and thus less dense, and have a higher S\'ersic parameter than field galaxies of similar mass. Similar findings have been made for other $z\sim2$ (proto-)clusters \citep{papovich12,strazzullo13}, while no differences in the structural properties of massive quiescent galaxies were found in a massive cluster at $z=1.80$ \citep{newman14}. \citet{newman14} conclude that the combined evidence from all (proto-)cluster data combined is rather weak, but point out that difficulties arising from systematic differences in measurement techniques and working with non-spectroscopic samples do not rule out that some real structural differences already exist at this early epoch. Simulations of protocluster regions show that such environments may also be expected to contain massive galaxies that are more compact (denser) compared to those in the field, and that such galaxies require significant ``puffing up" in order to comply with local structural relations for massive galaxies \citep{sommer-larsen10}. \citet{wang16} found that both quiescent and star-forming massive members in the core of protocluster CL J1001 at $z=2.5$ are more compact compared to local cluster early-types, but no different compared to field galaxies at the same redshift. Thus, this may indicate that any of the changes that may occur to the sizes of galaxies in dense environments have not yet occurred.
\begin{figure}[t]
\includegraphics[width=\textwidth]{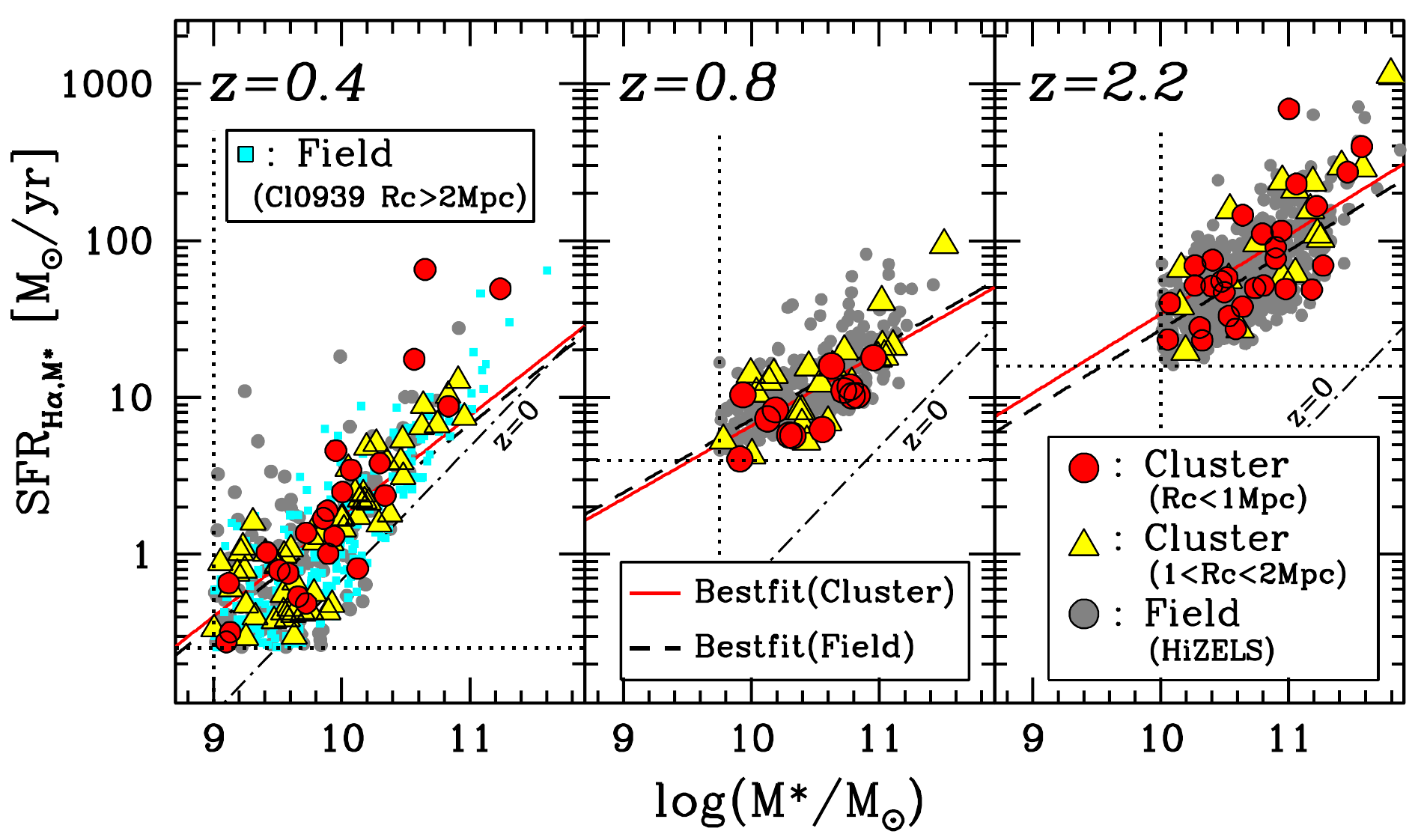}
\caption{\label{fig:koyama13b}The relation between stellar mass and \ha-based SFR for galaxies in (proto-)clusters and the field at three different redshifts: $z=0.4$ ({\it left panel}), $z=0.8$ ({\it middle panel}) and $z=2.2$ ({\it right panel}). Although significant evolution in the so-called main sequence of star formation with respect to the $z=0$ relation ({\it dot-dashed lines}) is seen for all three redshifts, the relation for field and (proto-)cluster regions is indistinguishable, except for a possible enhancement in the number of galaxies at the high mass end in (proto-)clusters. Figure reproduced from Fig. 8 in \citet{koyama13b}.}
\end{figure}

\subsubsection{Stellar populations and star formation} 

We have already discussed the available evidence on red sequence galaxies in protoclusters in Sect. \ref{sec:redsequence}. Most notably perhaps were the excesses of massive, quiescent galaxies based on spectroscopic data \citep[e.g.,][]{tanaka13,lemaux14}. \citet{steidel05} found that the star-forming galaxies in the protocluster HS1700+643 at $z=2.30$ are both more massive and older than an unbiased sample of field galaxies selected from neighboring redshifts (Fig. \ref{fig:steidel05}). Although this protocluster is not known to contain a cluster red sequence, these results are among the strongest indications that the properties of galaxies in some protoclusters differ strongly from those in the field. Based on a large sample of HAEs in the Spiderweb protocluster at $z=2.2$ \citep{koyama13a} and a control sample of HAEs in a large field survey, \citet{koyama13b} showed that the SFR-$M_*$ relation is indistinguishable between protocluster and field environments \citep[Fig. \ref{fig:koyama13b}; see also][]{cooke16}. However, a small excess of massive ($M_*>10^{11}$ $M_\odot$) galaxies relative to the field is seen, and these galaxies tend to also have higher SFRs relative to the field \citep[see also][]{kurk09,hatch11b}. \citet{hatch11b} also find that $z=2.2-2.4$ protoclusters (including the Spiderweb protocluster) contain HAEs that are, on average, twice as massive as HAEs in the field. They do not find any difference in their SFRs, but it is possible that the subtle effect reported by \citet{koyama13b} could only be seen due to their much larger HAE sample size. \citet{cooke14} furthermore found an excess of massive galaxies at $z=2.5$, while at the same time showing a rather puzzling deficit in the relative number of low-mass star-forming galaxies. 

The results summarized above shed some light on galaxy evolution during the epoch when the first differences between field and clusters started to appear. For completeness, however, we should also point out the rather large subset of protoclusters for which no differences are seen compared to the field, although these tend to be the higher redshift systems at $z\gtrsim4$ where the environmental effects are expected to be even smaller than at $z=2-3$ and more difficult to measure \citep[e.g.,][]{overzier08,toshikawa14,toshikawa16,kslee14}. In the cases where significant differences were found between protocluster and field galaxies, these differences are most naturally explained by taking into account that protocluster regions had an earlier start compared to the field. For example, \citet{steidel05} interpret the large age and mass difference as a consequence of assembly bias, given that the age difference between protocluster and field galaxies is close to the age difference one would naively expect on the basis of the overdensity and the earlier formation time of the galaxies forming within the overdensity. Results obtained in other protoclusters also argue in favor of an accelerated growth in overdense regions compared to the field \citep[e.g.,][]{hatch11b,matsuda11,koyama13b,cooke14}. However, we must remain somewhat cautious to compare results obtained with different field sizes or non-spectroscopic samples. Also, the possible contamination by AGN or improper dust corrections could easily skew the results towards or away from the true relations, as the sample of available protoclusters is still rather small. 

Despite the rather delicate evidence for environmental trends in terms of stellar populations in protoclusters discussed above, there is strong evidence that protoclusters are often associated with large excesses of dust-obscured star formation in the $z\simeq2-4$ redshift range \citep[e.g.,][]{stevens03,debreuck04,greve07,priddey08,daddi09,carrera11,ivison13,casey15,dole15,valtchanov13,dannerbauer14,rigby14,umehata14,umehata15} \citep[see][for an overview]{casey16}. The high levels of dust-obscured star formation in protoclusters suggests that they are rapidly forming massive galaxies. Protoclusters rich in dusty galaxies perhaps represent a particularly important phase during which a large fraction of the stellar mass on the present-day cluster red sequence was being formed during only a very brief time span in accordance with the small scatter of the cluster red sequence \citep[e.g.,][]{mei09}. The tail of this starburst phase can still be seen in some high redshift clusters which show unusual starbursts typically not seen in lower redshift clusters \citep[e.g.,][]{webb13,webb15a,webb15b,lee15,wang16}. As the starburst activity in (proto)clusters declines with decreasing redshift, the fraction of normal quiescent galaxies is seen to rise quite rapidly from $z\sim2.5$ to $z\sim1.5$ \citep[e.g.,][]{newman14,cooke15a,wang16}, indicating that the main epoch of massive cluster galaxy formation has ended. 
\begin{figure}[t]
\begin{center}
\includegraphics[width=0.9\textwidth]{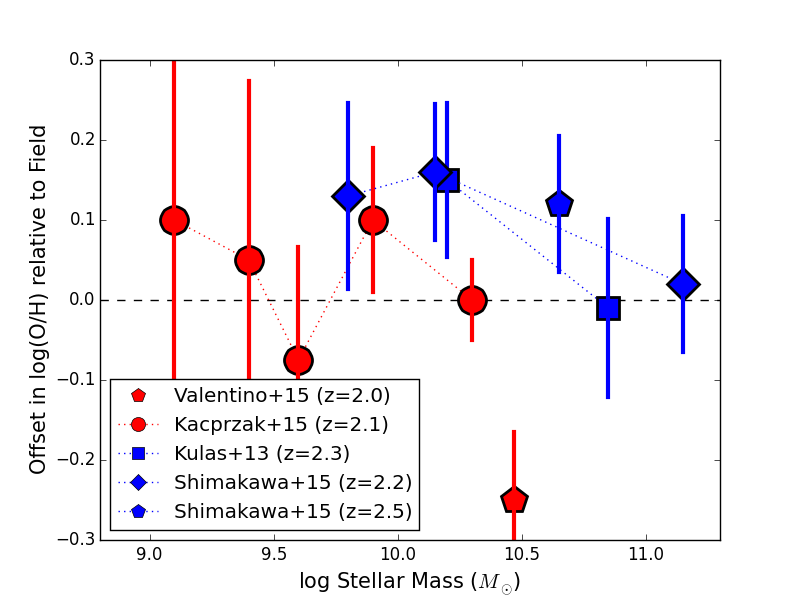}
\end{center}
\caption{\label{fig:metals}Compilation of stellar mass versus gas-phase oxygen abundance measurements of star-forming galaxies made in different environments in high redshift (proto)clusters. Measurements made at different stellar masses in the same systems are connected by dotted lines. The error bars on the data include both the statistical errors on the protocluster measurements and those on the field measurements. See the discussion in the text for details.}
\end{figure}

\subsubsection{Gas-phase metal abundances} 

Recent advances in instrumentation for performing deep multi-object spectroscopy in the near-infrared has allowed significant progress in the study of the nebular emission line properties of star-forming galaxies at high redshift. In Fig. \ref{fig:metals} we show a compilation of some of the first results related to the gas-phase oxygen abundances of galaxies in several high redshift (proto)clusters. Relatively low-mass ($M_*\sim10^{10}$ $M_\odot$) galaxies in three protocluster regions at $z\approx2.2$, $z\approx2.3$ and $z\approx2.5$ have marginally higher mean gas-phase metal abundances compared to field galaxies at the same redshift, while no such difference is seen for high mass ($M_*\sim10^{11}$ $M_\odot$) galaxies \citep{kulas13,shimakawa15}. 
In contrast, other studies have either found no differences between galaxies in (proto-)clusters and in the field \citep{kewley15,kacprzak15,tran15} or lower abundances in protoclusters \citep{valentino15}. Of course, it is still difficult to synthesize a consistent story based on these preliminary results on just a handful of systems. However, detailed investigation of the abundances of galaxies as a function of stellar mass and redshift in dense (proto)cluster environments and the field will be of great value to galaxy formation models. For example, enhancements in the metallicities of galaxies in protoclusters may be expected if these galaxies are accreting their gas from an IGM that is already metal-enhanced compared to lower density regions. The mass-dependent metallicity offset detected by \citet{kulas13} and \citet{shimakawa15} could then be the result of the more efficient recycling of enriched gas for low-mass galaxies in overdense regions which enhances the recycling rate of outflowing metal-enriched gas, as suggested by numerical simulations \citep{dave11,oppenheimer08}. The mean metallicity could also be increased due to stripping of the outer, more metal-poor gas through merging or ram-pressure stripping which should be more efficient in dense regions \citep{shimakawa15,kacprzak15}, while a dilution of the metallicity could also be expected due to the efficient inflow of gas from the IGM in the form of cold streams or mergers \citep{giavalisco11,cucciati14,valentino15}. 

\subsubsection{AGN} 

Several studies have found evidence for an elevated AGN fraction among protocluster galaxies \citep[e.g.,][]{pentericci02,lehmer09,digby-north10,lehmer13,chiang15}. Such increased AGN activity could be caused by the enhanced inflow rates of gas onto galaxies in overdense regions, either from the IGM or through galaxy merging. In addition, an elevated fraction of AGN is expected if protoclusters contain more massive galaxies relative to the field, given that the AGN fraction rises strongly as a function of stellar mass. The increased AGN fraction in protoclusters could be a reversal of the trend seen in groups and clusters at lower redshift, that tend to have AGN fractions that are either lower or equal to that in the field at $z\sim0$ and $z\sim1$, respectively \citep{martini13}. 

\section{Connections with \lya\ blobs, radio galaxies, QSOs, and reionization}
\label{sec:agn}

\subsection{Protoclusters and \lya\ blobs}
\label{sec:labs}

\begin{figure}[t]
\begin{center}
\includegraphics[height=0.45\textwidth]{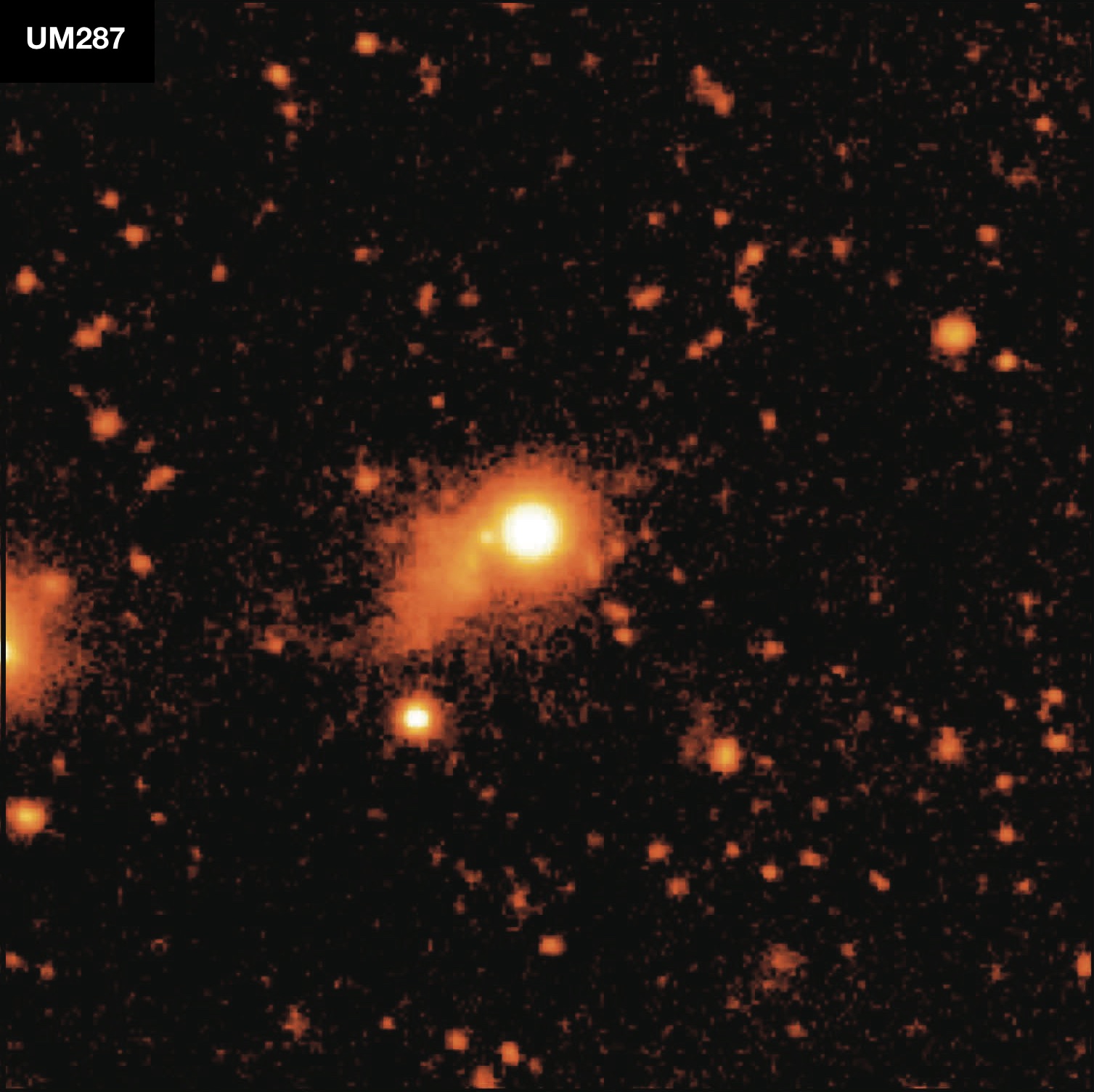}~~~
\includegraphics[height=0.45\textwidth]{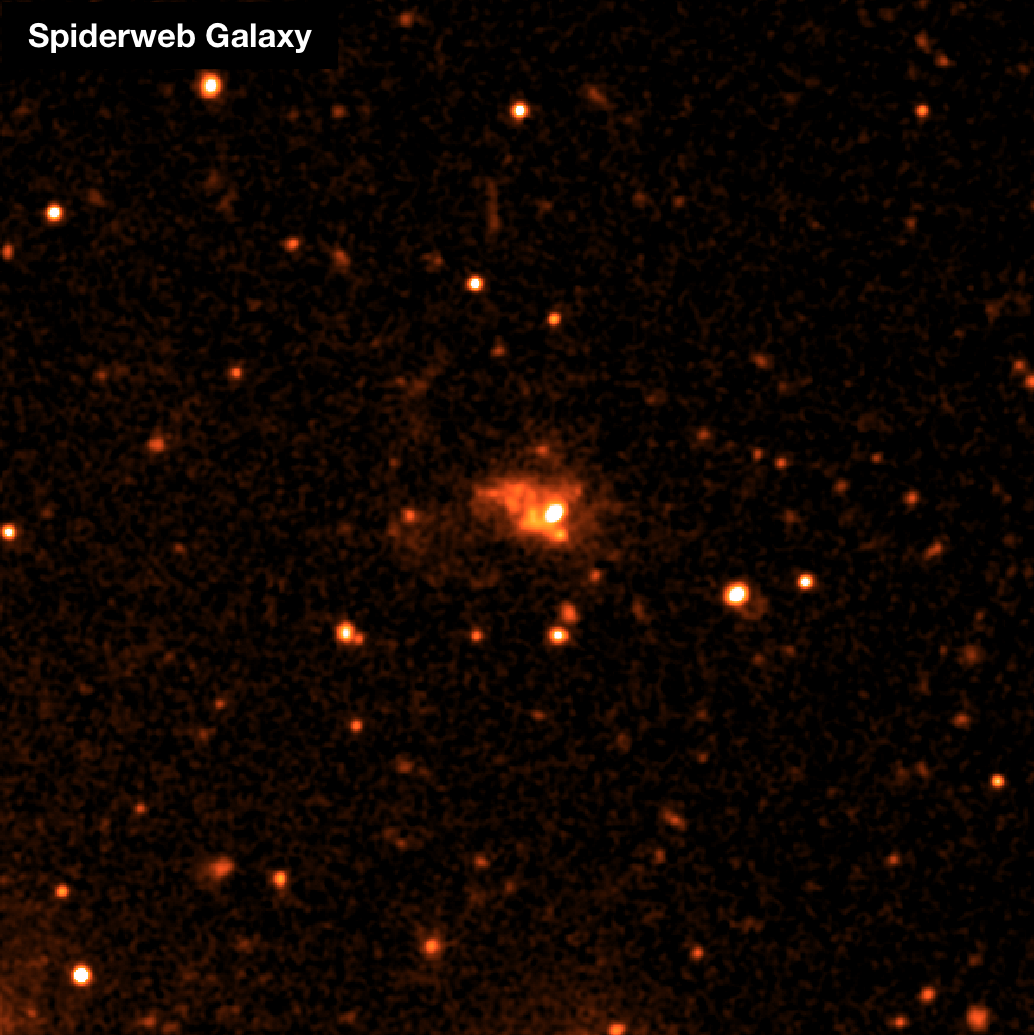}
\end{center}
\caption{\label{fig:labs}Extended \lya\ emission nebulae. {\it Panels} show narrow-band \lya\ images of the radio-quiet QSO UM287 at $z\approx2.3$ [{\it left panel} figure reproduced from Fig. 1a in \citet{cantalupo14}] and the Spiderweb radio galaxy at $z=2.2$ [{\it right panel} \citet{kurk00}]. Both images have a diameter of 2\arcmin, and the AGN are located at the center of {\it each panel}.}
\end{figure}

The enigmatic class of giant emission line nebulae commonly referred to as \lya\ halos (LAHs; when seen around radio galaxies and QSOs) and \lya\ blobs (LABs; in other cases) has been closely intertwined with the subject of protoclusters since their first discovery \citep[e.g.,][]{francis96,lefevre96,steidel00,venemans02,matsuda04}. %
These nebulae emit extremely luminous ($L_{\mathrm{Ly}\alpha}\sim10^{43-45}$ erg s$^{-1}$) \lya\ emission that is extended over tens or sometimes hundreds of kpc (see Fig. \ref{fig:labs}). They frequently occur around radio galaxies \citep[e.g.,][]{heckman82,baum89,mccarthy93,vanojik96,debreuck00,overzier01,venemans02,villar-martin07a}, radio-loud QSOs \citep[e.g.,][]{heckman91,borisova16}, radio-quiet QSOs and obscured AGN \citep[e.g.,][]{basu-zych04,chapman04,christensen06,dey05,geach09,yang09,fu09,overzier13,bridge13,cantalupo14,hennawi15,borisova16}, star-forming galaxies \citep[e.g.,][]{ouchi09,colbert11}, and sometimes without a clearly identifiable source \citep[e.g.,][]{steidel00,prescott08,erb11}. Extended \lya\ emission in galaxies having more typical luminosities of $L_{Ly\alpha}\sim10^{42-43}$ erg s$^{-1}$ (also referred to as LABs by some authors) is most likely due to scattered \lya\ photons from central or extended star formation, AGN, and perhaps some amount of gravitational cooling radiation \citep[e.g.,][]{hayashino04,saito06,zheng11,steidel11,dijkstra12,matsuda12,momose14,momose15,caminha15,patricio16}. In the case of the nebulae around radio galaxies and QSOs, there is little doubt that the powerful AGN are the main driver of the \lya\ phenomenon. This appears to be true for the majority of all other LABs as well (at least at high luminosities, $L_{\mathrm{Ly}\alpha}\gtrsim10^{44}$ erg s$^{-1}$), given that most LABs host a luminous obscured AGN \citep[see the overview in][]{overzier13}. Although luminous high redshift AGN are often surrounded by powerful starbursts that can provide additional ionizing photons \citep[e.g.,][]{stevens03,villar-martin07a,hatch08}, even in these sources the dominant power source of their \lya\ emission is the photoionization by the central AGN. Cases of LABs that are purely powered by star formation or gravitational cooling radiation must be very rare, otherwise many non-AGN LABs would be detected, especially considering that star formation and cooling radiation have significantly longer duty-cycles compared to AGN. 

It is intriguing that many protoclusters and overdense regions at high redshift known to date also host halos or LABs. In the case of protoclusters found near powerful high redshift radio galaxies this is obvious, given that these large extended emission line regions are conspicuous among these galaxies \citep{mccarthy93}. However, LABs have been found also in other overdense regions not associated with radio galaxies \citep[e.g.,][]{francis01,steidel00,matsuda04,matsuda09,saito06,prescott08,prescott12,erb11,hayashi12,cantalupo14,hennawi15,chiang15}, suggesting that the connection between protoclusters and large \lya\ nebulae could be more fundamental. The fact that the largest and most luminous halos and LABs are nearly always associated with AGN (radio-loud and radio-quiet obscured AGN and QSOs) points to a relation between protoclusters and (powerful) AGN. This connection can be understood, at least at the qualitative level. AGN are powered by the SMBHs in massive galaxies. The formation of these SMBHs relies on the merging between galaxies, while the triggering of the AGN requires a large supply of cold gas in their nuclear region. This gas could be supplied by mergers, or through infall from the circumgalactic medium. Overdense environments at high redshift are inducive to frequent mergers between (massive) galaxies, and are expected to be associated with high densities of cold gas recently accreted from the IGM. 

Besides the triggering of the AGN, we also need to explain the presence of the gas on scales of tens or hundreds of kpc around the AGN in the cores of protoclusters. \lya\ halos and LABs frequently show highly perturbed or outflowing gas in their inner parts (and  along radio jets, if present), while infall or rotating gas dominates the outer parts \citep[e.g.,][]{vanojik96,nesvadba06,villar-martin07b,humphrey08,martin14,martin15,prescott15}. The extended gas reservoirs around radio galaxies such as the Spiderweb Galaxy could perhaps in large part be explained by strong mechanical feedback from the radio jets \citep{nesvadba06}. However, the fact that perturbed gas occurs both in radio-loud and radio-quiet sources indicates that feedback does not necessarily require radio jets, but can also arise from QSO or starburst winds \citep{polletta11,liu13,overzier13,hennawi15}. In this case, it is unclear whether the presence of gas found at large distances from QSOs (and some radio galaxies) is consistent with the outflow scenario \citep[e.g.,][]{prochaska09,cantalupo14,johnson15,hennawi15}. Quasar hosts are massive galaxies that live in massive dark matter halos ($\sim10^{12-13}$ $M_\odot$) that should already be filled with substantial amounts of shock-heated gas of temperatures of $\sim10^7$ K at $z\sim2$ \citep[e.g.,][see also Fig. \ref{fig:mydekel}]{birnboim03,dekel06,ocvirk08,cattaneo08,fumagalli14}. A possible scenario is thus that the \lya\ emitting clouds form within the hot plasma in the QSO halos \citep{momiralda96}. In some systems observed, the QSOs appear to illuminate gas in the IGM, possibly the accretion flows, given that the physical extent of some nebulae exceed the virial radius of their dark matter halos \citep{cantalupo14,martin15,borisova16}. Another possibility is that the \lya\ emission arises from gas associated with the debris of massive mergers that is being illuminated by the AGN \citep{yajima13,yang14,hennawi15,johnson15}. \lya\ halos and LABs often show substantial chemical enrichment \citep[][]{debreuck00,overzier01,prochaska09,hennawi15}, which would perhaps be most easily explained by the outflow or merger scenario, rather than a scenario where the gas originates from the IGM. 

In addition to the luminous \lya\ systems discussed here, \citet{matsuda12} showed that also the relatively small and faint \lya\ emission halos seen around ordinary star forming galaxies may contain important information about the detailed structure of protoclusters. They suggest that in overdense regions the strong clustering between dark matter halos may lead to a different cold gas distribution that increases the scale-length of \lya\ \citep{zheng11}. In the future, deep \lya\ observations of protoclusters may thus illuminate the infalling gas on large scales, the in- and outflowing gas close to the central sources, and map the small-scale gas distribution between galaxies in overdense regions (see Sect. \ref{sec:finalremarks}). 

\subsection{Protoclusters associated with radio galaxies and QSOs}
\label{sec:rgqso}

We have discussed some of the evidence for protoclusters that have been found by targeting radio galaxies and QSOs, but also seen evidence of spectacular examples of protoclusters that do not, or at least are not known to, contain any particularly powerful AGN. This raises a number of interesting questions. In Sect. \ref{sec:rgqso2} we will first discuss the protocluster environments of radio galaxies and QSOs in general. In Sect. \ref{sec:z6qsos} we will look more closely at the environments of QSOs at $z\sim6$ given the strong interest in this area, specifically. 

\subsubsection{The environments of radio galaxies and QSOs}
\label{sec:rgqso2}

Based on the literature overview presented in Sect. \ref{sec:zoo}, a significant fraction of known protoclusters was found by targeting radio galaxies, suggesting a deeper connection between the two. In order to get some idea of the statistics on the occurrence of protoclusters near radio galaxies, it is best to look at studies that targeted large samples. Unfortunately, there are only few of such studies to date. \citet{venemans07} found that at least 75 \% of radio galaxies are in protoclusters traced by overdensities of LAEs, based on their sample of 9 radio galaxies ($z\approx2-5$ and $L_{2.7GHz}>10^{33}$ erg s$^{-1}$ Hz$^{-1}$). The evidence for protoclusters around these systems was primarily based on large overdensities of spectroscopically confirmed LAEs, but has since been corroborated by other galaxy populations in a number of cases. More evidence for protoclusters around radio galaxies has been uncovered since the seminal work of \citet{venemans07}, such as some newly identified protoclusters around other radio galaxies \citep{matsuda09,hatch11b,hayashi12,cooke14,shimakawa14} and a number of much larger statistical studies \citep[e.g.,][]{hatch11a,galametz12,wylezalek13}. 

\citet{hatch11a} performed a near-IR selection of galaxies in six radio galaxy fields at $z\sim2.4$, finding three fields to be overdense by a factor of $\sim3$ ($3\sigma$ deviations from the field surface density on scales of $R<3$ comoving Mpc), consistent with protoclusters \citep{hatch14}. Although the study used color selections that have considerable redshift uncertainty, one of the three fields was later shown to be overdense in narrow-band selected HAEs as well \citep{cooke14}. \citet{galametz12} observed 48 radio galaxies at $1.2<z<3$ using Spitzer and measured the surface densities of galaxies selected in the mid-IR. They found that 73 \% of their fields was denser than the median field survey, while 23 \% was denser than the field at the $\gtrsim2\sigma$ level. \citet{galametz12} and \citet{hatch14} also analyzed trends between overdensity and redshift or radio luminosity, finding no correlations. The CARLA survey targeted 387 radio-loud AGN at $1.3<z<3.2$ \citep[187 radio-loud QSOs and 200 radio galaxies;][]{wylezalek13}. The results indicate that 55 \% (10 \%) of the AGN fields are overdense at the $2\sigma$ ($5\sigma$) level.  No difference was found between the environments of the radio galaxies and the radio-loud QSOs, consistent with the idea that orientation is the only distinguishing parameter that separates these two populations. If we optimistically assume that about half of the radio AGN studied by \citet{wylezalek13} indeed lie in $2-5\sigma$ overdensities, this would imply that the other half of the sample occupies much more average and even underdense regions. However, it is not clear whether the fields in which no or only small overdensities were detected indeed do not contain protoclusters. The study recovered several previously known overdensities (clusters and protoclusters), while their color-selection also proved relatively insensitive to even some known protoclusters in these fields. Another complication arises from the fact that the tracer galaxies (e.g., radio galaxies or QSOs) are usually assumed to lie at or near the center of their (proto-)cluster. Many of the fields show substantial sub-clustering, and fields that do not show strong central overdensities close to the radio source often still show an excess just a few arcminutes away \citep{hatch14}. Although these studies were the first to measure the environments for a very large number of high redshift radio sources statistically, they naturally lack the redshift precision of, e.g., \citet{venemans07} due to the limitations inherent to the mid-IR color selection \citep{falder11}. A definite answer to the question what fraction of radio sources are in (proto-)clusters must therefore await future work. 

Many searches for protoclusters by specifically targeting QSOs have also been performed \citep[e.g.,][]{campos99,djorgovski03,wold03,kashikawa07,falder11,husband13,adams15,hennawi15}. A large statistical survey very similar to the Spitzer study of \citet{wylezalek13} was performed by \citet{falder11}. They analyzed the environments of 46 luminous ($M_i\le-26$), predominantly radio-quiet, QSOs. They also found significant overdensities based on the surface densities of sources around QSO subsamples centered at $z\sim2$ ($>4\sigma$) and $z\sim3.3$ ($>2\sigma$), interpreted as evidence for protoclusters similar to \citet{wylezalek13}. However, several recent works based on spectroscopic observations of LBGs and LAEs more similar to the \citet{venemans07} work on radio galaxy environments, have found evidence for protoclusters near luminous QSOs in about 10 \% of the cases \citep{trainor12,adams15,hennawi15}, although in at least one case the overdensity is one of the largest known \citep{hennawi15}. These results suggest that the success rate of protocluster discovery may be lower when using QSOs compared to radio galaxies. 

How can we explain this apparent discrepancy between the (protocluster) environments of radio galaxies and (radio-quiet) QSOs? Luminous QSOs and radio galaxies at high redshift are strongly clustered sources in average to massive dark matter halos \citep[e.g.,][]{shen07,shen09,overzier03,donoso10,falder11,white12,eftekharzadeh15}. At $z\sim2-3$, typical optical QSOs live in halos of $M_\mathrm{h}\sim2\times10^{12}$ $h^{-1}$ $M_\odot$ \citep[][]{white12,trainor12,eftekharzadeh15}, with no measurable dependence on redshift or luminosity. The halos of these QSOs will evolve into halos of $\sim10^{13}$ $h^{-1}$ $M_\odot$ today, which is an order of magnitude lower than those expected if these QSOs traced protoclusters at $z\sim3$ \citep[][and Fig. \ref{fig:pcsizes}]{chiang13}. Because of short duty cycles and the lack of correlation between luminosity and clustering, QSOs in general tend to trace average halos, regardless of luminosity or black hole mass \citep{font-ribera13,eftekharzadeh15,orsi15}. This explains, at least in part, the results of the many different QSO environment studies. However, \citet{shen07} found that the most luminous QSOs at $3<z<4.5$ have a much larger correlation length that is comparable to that of massive clusters today \citep{postman92}. The (minimum and median) masses inferred are an order of magnitude higher than those of the earlier sample ($M_h\sim4-6\times10^{12}$ $h^{-1}$ $M_\odot$ at $z\sim4$), and in the range of those expected for the central halos of protoclusters at these redshifts (Fig. \ref{fig:pcsizes}). 
As argued by \citet{adams15}, these QSOs should have a high duty cycle ($\gtrsim10$ \%) and lie on a relatively tight relation between QSO luminosity and halo mass \citep[see][]{martini01,shankar10}, in contrast to the more typical optical QSOs at these and lower redshifts \citep[see also][]{mcgreer16}. Radio-loud QSOs at $z\lesssim2$ are furthermore more strongly clustered than radio-quiet QSOs matched in luminosity or SMBH mass \citep[$M_h\sim10^{13}$ $h^{-1}$ $M_\odot$;][]{shen09}, and radio galaxies at $z\sim1.5$ have $M_h\sim10^{13.6}$ $M_\odot$ \citep{allison15}. If this correlation between characteristic halo mass and radio-loudness extends to higher redshifts, we would expect that protoclusters are indeed found more frequently near radio galaxies relative to QSOs. 

These results are consistent with predictions from semi-analytical models \citep[e.g.,][]{fanidakis11,fanidakis13,orsi15}. Radio galaxies are typically hosted by the most massive haloes present at any redshift, and their typical $z=0$ descendant halo is more massive than that of the halos hosting typical QSOs. In these models, the difference between radio galaxy and QSO hosts mainly comes about because the radio power is assumed to scale with black hole mass and spin, both of which are larger in more massive halos. Radio galaxies furthermore rely on a low accretion rate regime to allow the formation of an advection dominated accretion flow that fosters radio jets, while QSOs are assumed to form in high accretion rate regions that do not lead to strong jets \citep[e.g., see][]{fanidakis11,fanidakis13,hatch14,orsi15}. Radio sources at $z>1$ are found in denser-than-average structures compared to ordinary galaxies of the same stellar mass, indicating that radio activity may indeed be preferentially triggered in dense environments \citep[see also][]{rawlings04}. The triggering of radio jets requires both the presence of a rapidly spinning SMBH with $M_\mathrm{BH}\gtrsim10^8$ $M_\odot$ \citep{mclure04,fanidakis11} as well as the inflow of gas. \citet{hatch14} suggest that these requirements are most easily met in the gas-rich cores of protoclusters where frequent mergers between massive galaxies furthermore help to increase the mass and spin of the SMBHs \citep[see also][]{chiaberge15}.

\subsubsection{The environments of QSOs at $z\sim6$}
\label{sec:z6qsos}

The luminous QSOs at $z\sim6$ are believed to be associated with dense matter peaks in the early universe. Their very low space density of about 1 per Gpc$^3$ implies a (maximum) host halo mass of $\sim10^{13}$ $M_\odot$ \citep[e.g.,][]{fan01,springel05,li07}. A halo that is that massive at $z\sim6$ will easily evolve into a cluster-sized halo by the present-day, even if it experiences only modest (say, a factor of 10) growth. Based on the expectation that these extreme high redshift objects should pinpoint the locations where massive clusters of galaxies are forming, a significant amount of attention has recently been devoted to studying the environments of luminous QSOs at $z\sim6$. 

To date, the environments of $\approx$13 QSOs between $z=5.7$ and $z=7.1$ have been investigated, searching for faint LBGs in deep pencil-beam surveys with HST \citep[][]{stiavelli05,zheng06,kim09,mcgreer14}, or for bright LBGs \citep[][]{willott05,utsumi10,simpson14,morselli14} and LAEs \citep{banados13,mazzucchelli16} in wider field ground-based data \citep[see][for an overview]{mazzucchelli16}. 
\citet{willott05} found no evidence for excesses of $z^\prime<25.5$ LBGs within $\sim10$ Mpc (co-moving) fields around 3 QSOs, while \citet{utsumi10} and \citet{morselli14} found excesses within much wider $\sim$50 (co-moving) Mpc fields around 5 QSOs [including those of \citet{willott05}]. Some QSOs show an excess of fainter LBGs detected with HST \citep{stiavelli05,zheng06}, but these appear to be not as common as QSOs in relatively underdense or average density regions \citep{kim09,mcgreer14,simpson14}. Perhaps the strongest constraints on QSO environments at these redshifts come from narrow-band observations, which did not find any excess of LAEs within $\Delta z\approx0.1$ for 2 QSOs \citep{banados13,mazzucchelli16}.   

\begin{figure}[t]
\begin{center}
\includegraphics[width=0.75\textwidth]{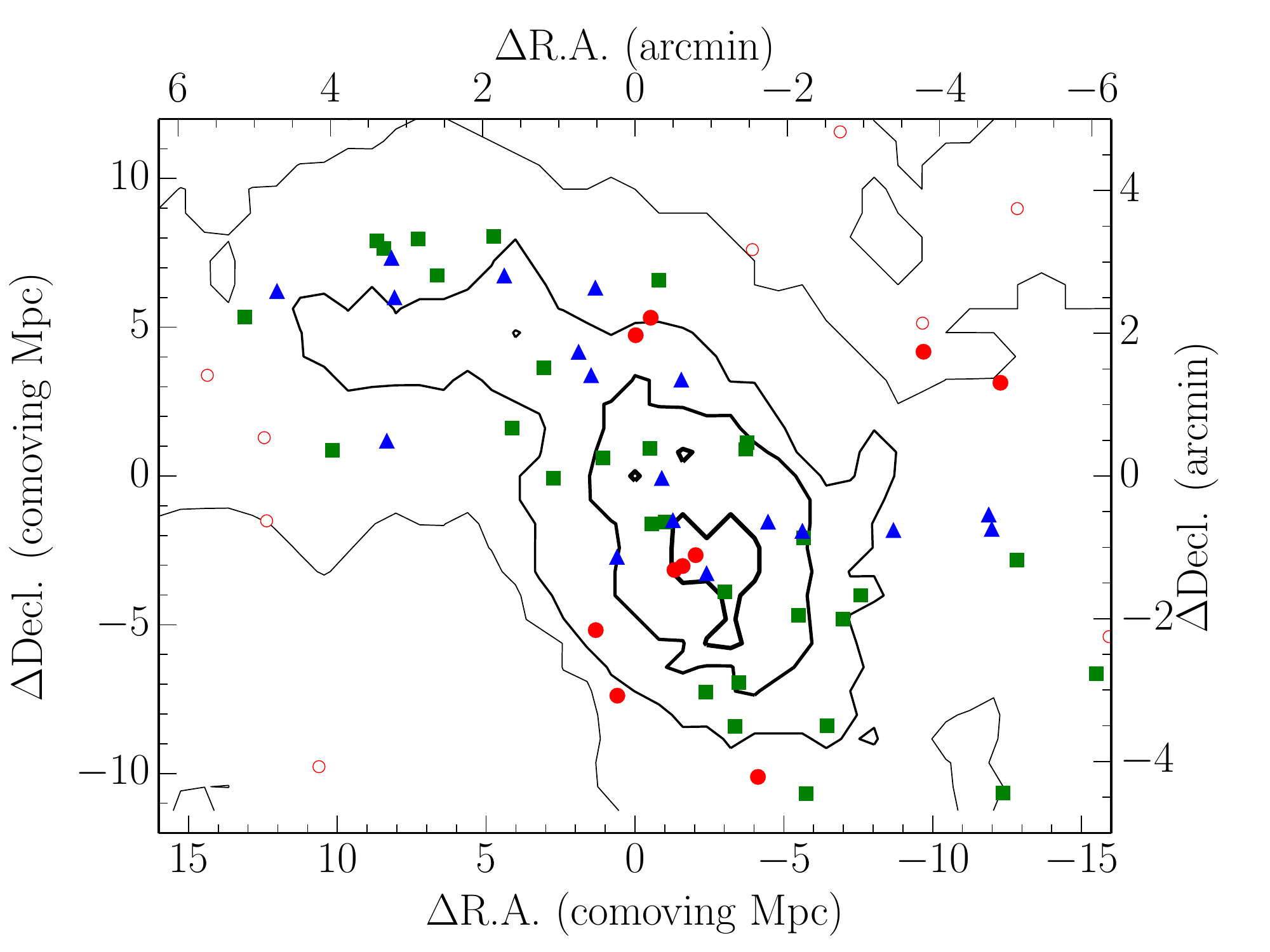}
\end{center}
\caption{\label{fig:toshikawa14}A protocluster at $z\approx6.01$ in the Subaru Deep Field \citep[SDF;][]{toshikawa12,toshikawa14}. {\it Red solid circles} mark the positions of the 10 spectroscopically confirmed protocluster members, while blue triangles mark other $z\sim6$ galaxies. $i$-dropouts without spectroscopic confirmation of \lya\ are indicated by the {\it green squares}. Contours indicate the surface density of $i$-dropouts at 0--6$\sigma$. The large excess and significant clustering of \lya\ and dropout galaxies in the SDF protocluster are stronger than that observed near any of the $z\sim6$ QSOs studied to date. Figure reproduced from Fig. 1 in \citet{toshikawa14}.}
\end{figure}

Summarizing these results, there is currently very little observational evidence that supports the conclusion that the $z\sim6$ QSOs are in large overdense regions or protoclusters. We note that many studies relied on the dropout technique to isolate $z\sim6$ galaxies in the vicinity of the QSOs. Although this technique works reliably for such galaxies in the field, it is sensitive to galaxies from a wide redshift interval ($\Delta z\sim0.6$). If the QSOs typically trace large overdense regions, we may still have expected to see a statistical excess of dropout galaxies despite the relatively crude redshift selection, but this does not appear to be the case for most fields observed. On the other hand, for QSO fields that are average or slightly over- or underdense, we cannot rule out that deeper data or spectroscopic observations may still reveal significant overdensities \citep[see][for such cases at $z\sim5$]{husband13}. Although a full census of the QSO environments will need to await future studies, we can, however, already state with certainty that there exist highly overdense regions at $z\sim6$ that display much larger concentrations of galaxies than those observed near QSOs \citep[e.g.,][]{ouchi05,toshikawa14,ishigaki16}. For example, the $z=6.01$ protocluster studied by \citet{toshikawa12,toshikawa14} is one of the largest concentrations of galaxies found at these redshifts, and such a structure would have easily been detected if it existed around any of the $z\sim6$ QSOs previously studied (Fig. \ref{fig:toshikawa14}). 

Both the observations described above and simulations \citep{overzier09a,romano-diaz11} suggest that the environments of QSOs observed are inconsistent with these sources tracing the richest regions of the early universe, in stark contrast with the simple theoretical expectations that motivated many of these studies. It is unclear what causes this discrepancy. In principle, it is still possible that current observations failed to detect significant overdensities because they do not reach a sufficient depth or area, or because the galaxies in the QSO environment are highly obscured by dust. However, this would be somewhat surprising given that several highly significant (non-QSO) protoclusters were found at $z\sim6$ using data of very similar quality as that obtained for the QSO fields. Also, no dusty star-forming galaxies have been found in early Atacama Large Millimeter Array (ALMA) observations of $z>6$ QSOs \citep{venemans16}. Other authors have suggested that strong radiative feedback may suppress galaxy formation in QSO environments \citep[e.g.,][]{kashikawa07,utsumi10}, but it is not clear if QSO feedback could offset gas cooling uniformly in all directions and on scales as large as the full extent of protocluster regions (several tens of Mpc at $z\sim6$). An alternative option that is consistent with most of the data, is that the $z\sim6$ QSOs may trace regions that, on average, are significantly less overdense compared to the theoretical expectation. 

Simulations show that there is a non-trivial relation between the halo mass at $z\sim6$ and that of its descendent at $z=0$ \citep[e.g.,][]{overzier09a,romano-diaz11,angulo12}. \citet{angulo12} showed that there is a good linear correlation between the overdensity at $z=6$ and the halo mass at $z=0$ that is stronger than the correlation between halo mass at the two redshifts. Furthermore, the slope of the relation between overdensity and $z=0$ mass is steeper and has a smaller scatter when the density is determined on large smoothing scales ($\sim$7--28 Mpc) compared to small smoothing scales ($<$ few Mpc). This suggests that it is the large-scale environment of a $z\sim6$ QSO that determines whether a massive cluster will develop or not. These results may very well explain the observations of QSO environments at these redshifts. Some important guidelines that can be deduced are the following. The redshift evolution of massive dark matter halos between $z\approx6$ and $z\approx0$ is not as straightforward as often assumed, and the most massive halo at $z\sim6$ will typically not be the most massive halo at $z=0$ \citep{angulo12}. This could be the case for some of the early protoclusters detected \citep{ouchi05,toshikawa14,ishigaki16}. Also, there is a non-zero possibility that QSO-hosting halos of relatively low mass at $z\sim6$ will still become among the most massive halos by $z=0$. In this scenario, the QSOs could be in average or mildly overdense environments at $z\sim6$ and still end up near the top of the local $M_\bullet-\sigma_*$ relation \citep{mcconnell12}. In any case, we expect a large scatter between the overdensity of halos measured at $z\sim6$ and the present-day descendant halo mass. Overdensities measured on very large ($\gtrsim7$ Mpc) scales are a better indicator than those measured on small scales \citep{angulo12}, and it will thus be important to obtain sufficiently deep observations over very large fields of view before we can rule out that $z\sim6$ QSOs end up in massive clusters today. 

It will be interesting to see a definitive conclusion about the extreme (or not) environments of the first luminous QSOs at $z\gtrsim6$. Deep narrow-band studies are limited to a number of atmospheric windows where the OH night-sky emission is low. \citet{banados13} therefore suggest to specifically search for QSOs at $z\approx5.7$, $z\approx6.6$ and $z\approx7.0$, and then target their environments in the narrow redshift intervals of the filter. In light of the \citet{angulo12} predictions, it will be important to search for LBGs and LAEs in very wide fields around these QSOs, while ALMA could be used to search for the far-infrared line emission from obscured galaxies at the redshifts of the QSOs. 

\subsection{Protoclusters and the epoch of reionization}
\label{sec:reionization}

The epoch of reionization (EoR), during which hydrogen in the universe transitioned from being mostly neutral to being mostly ionized, is an extremely important event for cosmology, structure formation and galaxy evolution \citep[e.g., see][for a review]{mellema13}. Although in principle there are a number of candidates for the sources responsible for reionization, such as individual stars, galaxies and (mini-)quasars, it seems likely that the stars of the first generation of galaxies were responsible for producing the bulk of the Lyman continuum photons that ionized the neutral IGM \citep[e.g., see][]{robertson15,bouwens15b}. 

\begin{figure}[t]
\begin{center}
\includegraphics[width=0.8\textwidth]{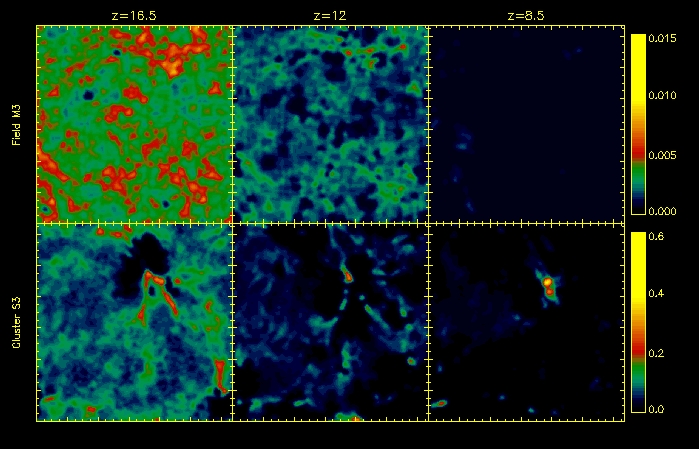}
\end{center}
\caption{\label{fig:ciardi03}Simulated maps of the neutral and ionized IGM during the EoR in field ({\it top panels}) and protocluster regions ({\it bottom panels}). Figure reproduced from Fig. 4 in \citet{ciardi03}.}
\end{figure}

Simulations of structure formation show that as early as $z>6$, the universe contains significant density fluctuations on large scales (tens of Mpc, see Fig. \ref{fig:boylankolchin09}). The early ionizing radiation field was dominated by the combined star formation of galaxies in these large overdense regions. Once the ionized bubbles around the large numbers of individual galaxies in these regions made contact, large ionized regions measuring tens of Mpc in diameter formed surrounded by a sea of neutral hydrogen. As reionization proceeded, the ionization front penetrated from the high into the surrounding lower density regions in the so-called ``inside-out'' fashion. It is thus expected that early density fluctuations will lead to a patchy reionization on large scales. An example showing the differences expected between field and protocluster regions is shown in Fig. \ref{fig:ciardi03}, reproduced here from \citet{ciardi03}. Initially, the high photon production rate in large overdensities causes completely ionized regions  around protoclusters that are both larger and that have an overall lower neutral fraction compared to the field at similar redshifts (left and middle panels of Fig. \ref{fig:ciardi03}). At late times, however, the cores of protoclusters may be among the last regions to reionize given the high recombination rates in these very dense regions (right panels). Large simulations performed by \citet{iliev14} also find that inside-out reionization leads to large voids that remain neutral longer compared to the high-density peaks and filaments. These differences will be reflected in the fluctuations in the redshifted 21 cm emission line signal used to probe this epoch. The redshift at which a certain ionized fraction ($f_\mathrm{HII}$) is reached can differ substantially depending on whether a region is under- or overdense with respect to the mean density. \citet{iliev14} find that for a fixed value of $f_\mathrm{HII}$, the redshifts at which those ionized fractions are obtained differ by as much as $\pm0.04$, $\pm0.2$ and $\pm0.5$ when measured on scales of $\sim$200, 100 and 50 Mpc. 
The mean redshift at which the typical $\sim50$ Mpc-size region reaches an ionized fraction of 10 \% (90 \%) is $z\approx8.6$ ($z\approx6.9$), whereas these redshifts are  $z\approx9.1$ ($z\approx7.3$) for regions that are 10 \% overdense compared to the mean density \citep[see Fig. 13 in ][]{iliev14}. 

\citet{cen05} have suggested that the ionizing photons produced by the progenitors of the faint dwarf galaxies in present-day clusters may have been responsible for a large fraction of the photons required for reionization. In fact, the study showed that protoclusters could be capable of fully ionizing the IGM entirely by themselves. Because the earliest HII regions will form in large overdense regions that are highly biased and strongly clustered, the overlapping of protocluster-sized regions will create large ionized regions and large spatial variations in maps of the neutral gas fraction. It is thus highly plausible that protoclusters of galaxies played an important, perhaps even dominant role in the process of reionization. Upcoming reionization measurements may be able to see these protoclusters, first as the very large-scale overdensities of neutral hydrogen before reionization starts, then as the first very large patches of relatively low neutral hydrogen fraction. In some sense, reionization studies will be a study of cluster formation. 

It is interesting to note that the very rare and brightest dropout galaxies seen toward the end of reionization near $z\sim6-10$ have number densities of $\lesssim10^{-6}$ Mpc$^{-3}$, i.e., equal to those of rich clusters today \citep{bouwens15c}. It would be interesting to know whether this is just a coincidence, or whether these highly biased galaxies can only exist because they formed as part of the large overdense regions associated with protoclusters. \citet{ono12} has found an indication that the fraction of \lya\ emitting galaxies drops more strongly between $z\sim7$ and $z\sim6$ in UV-faint galaxies compared to UV-bright galaxies. This suggests that the IGM, which is believed to be responsible for the absorption of \lya\ in star-forming galaxies, became more neutral toward higher redshift but the change is not as pronounced for brighter galaxies. This is consistent with the inside-out reionization model if the brighter LAEs trace higher density regions in which reionization was completed earlier compared to the lower density regions traced by the fainter galaxies. 

Besides these direct links between protoclusters and the EoR, there is yet another way in which protoclusters can play an important role. Lyman continuum observations of high redshift galaxies can give important clues to the processes responsible for the escape of ionizing photons. Protoclusters contain large numbers of star-forming galaxies that are relatively bright in the rest-frame far-UV (LBGs and LAEs). Because these galaxies have a narrow redshift distribution and occupy a relatively compact area on the sky, protoclusters are good targets for studying the escape of ionizing photons based on deep imaging or multi-object spectroscopic observations below the Lyman limit \citep[see][for an example using a protocluster at $z=2.85$]{mostardi13}.

\section{Outlook}
\label{sec:future}

\subsection{Future observations}
\label{sec:surveys}

We have seen in this review that there exist excellent examples of protoclusters over a wide range in redshift and with a variety of distinguishing properties that have been used to shed light on important aspects of cluster formation prior to $z\sim2$. However, if we want to observe this process from its earliest stages to the present, we will need to obtain statistical and unbiased samples of protoclusters in various bins of redshift, mass and evolutionary stage, rather than the highly heterogeneous and relatively limited data set we have available today. Fortunately, there are a number of new instruments and surveys that should soon be able to provide us for the first time with the large samples needed for such an analysis. Because protoclusters are rare and high redshift galaxies are faint, protocluster science should benefit enormously from deep, wide optical and near-infrared imaging surveys. Surveys with a high discovery potential are the Kilo-Degree Survey \citep[KiDS; 1500 deg$^2$ in $ugri$;][]{dejong13}, the Dark Energy Survey \citep[DES; 5000 deg$^2$ in $grizY$;][]{diehl14} and the Hyper Suprime-Cam Wide survey \citep[HSC-Wide; 1400 deg$^2$ in $grizY$;][]{takada14}. Due to its greater depth, HSC-Wide will be the most promising for finding large numbers of protoclusters based on the detection of surface overdensities of high redshift dropout galaxies. In a recent pilot project based on CFHTLS data, \citet{toshikawa16} found that about 80 \% of photometrically selected $\delta>4\sigma$ galaxy overdensities correspond to actual protoclusters of massive present-day clusters. The number density of such regions is about 1 deg$^{-2}$ per unit redshift from $z\sim3$ to $z\sim6$, which translates into more than 1000 high quality systems per unit redshift expected over the whole HSC-Wide area. Much smaller area (a few to a few tens of deg$^2$) but significantly deeper components of the HSC survey will include several narrow-band filters that will be used to perform a blind search for protoclusters using dropout galaxies and LAEs at specific, relatively narrow redshift intervals. The greater depth and efficient redshift selection will allow us to, e.g, compare UV and \lya\ luminosity functions in protoclusters and in the field, determine the \lya\ escape fraction as a function of overdensity, determine the topology and substructure of protocluster regions, search for correlations between overdensity, quasar activity and diffuse \lya\ emission (halos or LABs), or probe the distribution and metal abundance of the gas in the infall regions around protoclusters detected in absorption against background QSOs and galaxies. The Large Synoptic Survey Telescope (LSST) will probe an even larger area and to a greater depth compared to HSC-Wide (20000 deg$^2$ in $ugrizY$). Combining these deep optical surveys with matched observations in the 1--5 $\mu$m range will improve photometric redshifts and allow the selection of dusty star-forming galaxies and passively evolving galaxies at $z\simeq1-3$ and is therefore ideal to select and study high redshift (proto)cluster candidates and their stellar populations. The Wide Field Infrared Space Telescope (WFIRST) will perform large area imaging and spectroscopy in the near-IR to unprecedented depth that will enable us to both study the assembly of galaxy clusters and their galaxies at $z\simeq1-3$, and to obtain large statistical samples of galaxies, quasars and protoclusters during the important first billion years of cosmic history.  

Perhaps the highest potential for protocluster studies will be to sample the cosmic web at high redshift with high spectroscopic completeness. If the volumes are large enough, protoclusters will naturally appear in the catalogs extracted from these surveys without having to make any pre-selections or guesses about what constitutes a protocluster or where to find them. Also, such surveys will be able to immediately provide detailed physical properties related to the kinematics, stellar populations and gas properties of galaxies inside and outside the overdense regions, provided that the spectra have sufficient signal-to-noise and spectral resolution \citep[e.g.,][]{steidel00,steidel05,cucciati14,lemaux14}. New large multiplexed fiber-fed spectrographs that are being planned for 4--10 m telescopes will be extremely suitable for extending such studies to larger areas of the sky and a larger redshift range. 

For example, the Hobby--Eberly Telescope Dark Energy Experiment \citep[HETDEX;][]{hill08} is a blind 450 deg$^2$ spectroscopic survey sensitive to LAEs at $1.9 < z < 3.5$ that offers a strong advantage for finding homogeneous samples of protoclusters compared to standard broad- and narrowband imaging surveys \citep{chiang13,orsi15}. \citet{chiang15} showed that HETDEX data will be sufficient to estimate the $z=0$ cluster masses corresponding to the overdensities of LAEs in protoclusters. HETDEX will furthermore detect quasars, radio galaxies and LABs, providing relatively unbiased and statistically meaningful data on the relation between such objects and protoclusters. Other advances will come from the next generation of highly multiplexed fiber spectrographs. The William Herschel Telescope Enhanced Area Velocity Explorer \citep[WEAVE;][]{dalton12} and the 4 m Multi Object Spectroscopic Telescope \citep[4MOST;][]{dejong14} will both have of order 1000 fibers over a few square degree field of view to perform medium resolution (MR) spectroscopy. The Multi-Object Optical and Near-infrared Spectrograph \citep[MOONS;][]{cirasuolo14} is a new $\sim1000$ fiber MR spectrograph planned for the Very Large Telescope (VLT), while the Prime Focus Spectrograph \citep[PFS;][]{takada14} for the Subaru Telescope will be capable of performing wide, deep spectroscopic surveys owing to its $\sim$2400 fibers that can be positioned across a 1.3-deg field. Surveys with such large multiplexing capacity will be able to detect high redshift star-forming galaxies (mostly LAEs and LBGs) with sub-Mpc (comoving) line of sight resolutions. At such a high resolution the measured overdensities of large-scale structures such as protoclusters will be very close to their true spherical overdensities, unlike narrow-band surveys that typically probe comoving depths of $\sim100$ Mpc \citep{chiang13,orsi15}.  

Although the direct detection of 21 cm emission from individual galaxies with new radio facilities such as the Square Kilometer Array (SKA) and its pathfinders will be difficult at high redshift, protoclusters may still be discovered by searching for highly clustered star-forming galaxies and AGN detected spectroscopically through their lower-level CO transitions or, without redshifts, in the radio continuum. Very high redshift radio galaxies may be identified by pushing the ultra-steep spectrum technique that was used to discover many of the highest redshift radio sources known to date to even higher redshifts \citep{rottgering97,rottgering06}. Such distant $z>6$ radio sources would be very powerful tracers of protoclusters and SMBHs, and could also be used to study the IGM during the epoch of reionization by searching for 21 cm absorption along the line of sight \citep{ciardi13}.  

We have already seen in Sect. \ref{sec:igm} that IGM absorption techniques targeting background quasars and galaxies are extremely promising. The tomographic mapping of the cosmic web at high redshift relies on a dense packing of background sources, requiring high S/N spectra of star-forming galaxies rather than the much more luminous but much rarer QSOs. This requires 8m class telescopes and instruments such as MOONS and PFS. The IGM absorption studies will not only offer a new and competitive technique for finding galaxy protoclusters, they will also open up a new window to studying the local and bulk gas properties of these complex environments. 

In Sect. \ref{sec:reionization} we have shown that the mapping of the distribution of ionized and neutral hydrogen during the EoR in some sense also maps out the distribution and properties of protoclusters, given that a significant, perhaps even dominant, fraction of all star formation at these redshifts is taking place in these highly biased, overdense regions. Upcoming surveys that trace the neutral or ionized gas distribution during this epoch such as the SKA will be able to test this. 

It will remain challenging to increase significantly our capacity for detecting clusters of $M\gtrsim10^{14}$ $M_\odot$ at $z\gtrsim2$ even with state-of-the-art X-ray missions (e.g., eROSITA), SZ effect experiments (e.g., SPTpol, ACTpol), and optical/near-IR surveys (e.g., LSST, EUCLID) \citep[][]{weinberg13}. However, as shown in this review, protoclusters are typically identified through their characteristic large-scale overdensities of dark matter, galaxies or gas, and thus the technical limitations that affect classical cluster searches do not apply. A final argument for employing the enormous upcoming multi-wavelength survey capability is to provide the most suitable targets for detailed follow-up with facilities such as ALMA, the James Webb Space Telescope (JWST), and the planned 30-m class telescopes, as these facilities are not capable of performing large surveys.  

\subsection{Concluding remarks}
\label{sec:finalremarks}

The data assembled in this review demonstrate that the redshifts, sizes, galaxy and mass overdensities, collapse redshifts and descendant halo masses of many structures found in the $z\sim2-8$ redshift range are consistent with those expected for the progenitors of galaxy clusters (Sects. \ref{sec:observations} and \S\ \ref{sec:zoo}). However, we must remain skeptical both when discussing the properties of single systems as well as ensemble properties, as the measurements are still affected by small number statistics, substantial uncertainties, incompleteness and selection bias. 

\begin{figure}[t]
\begin{center}
\includegraphics[width=0.8\textwidth]{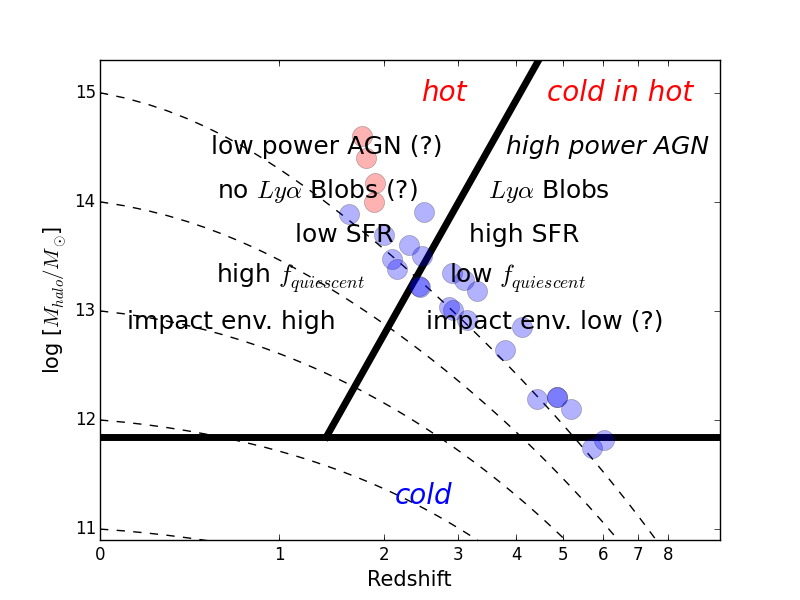}
\end{center}
\caption{\label{fig:dekel_trends} The growth history of protoclusters and clusters, highlighting some important evolutionary characteristics that can be tested with upcoming observations. The models and data shown are the same as in Fig. \ref{fig:mydekel} (see the caption of that figure for details).}
\end{figure}

The galaxies in these overdense regions will evolve from a protocluster-like regime that is  relatively gas-rich, rapidly star-forming and has only weak environmental impact, to a cluster-like regime that is relatively gas-poor, quiescent and has stronger environmental impact (Sect. \ref{sec:evolution}; Fig. \ref{fig:dekel_trends}). We have shown that the epoch between $z\sim2.5$ and $z\sim1.5$ should be a particularly important epoch for studying how exactly these transitions affect the properties of (proto)cluster galaxies and the ICM \citep[see also the discussion in][]{valentino15}, and protoclusters may thus be crucial for confirming several key aspects of galaxy and structure formation models. We have shown evidence that red sequence galaxies are starting to appear in some protoclusters, possibly at a rate that is faster than that in the field (Sect. \ref{sec:redsequence}). However, with the exception of \citet{lemaux14} and \citet{wang16}, few of these studies are based on spectroscopic evidence. New surveys such as VUDS are allowing for a more uniform selection of protocluster galaxies relatively independent of their star formation activity, although the spectroscopic confirmation still relies on the presence of emission lines or substantial flux in the UV part of the spectrum to identify breaks or absorption features. It is notoriously challenging to confirm spectroscopically truly ``passive'' galaxies at $z\gtrsim2$ \citep{kriek06,kriek09,toft12}, although this situation is improving \citep[e.g.,][]{kriek15}. Protoclusters also offer the unique opportunity to study the formation of BCGs (Sect. \ref{sec:bcg}). Although the observational data are still rather limited to only a few well-studied systems, the growing number of protoclusters will eventually allow us to track the growth of BCGs more consistently across the $z\simeq0-3$ redshift range in order to obtain an accurate picture of the SFHs and morphological evolution of the most massive galaxies in the universe and their SMBHs.  

The luminous extended \lya\ emission around radio galaxies, QSOs and LABs that is frequently seen in protoclusters appears to be related to the gas distribution around powerful AGN in massive overdense regions (Sect. \ref{sec:labs}). If so, their space density, luminosities, metal abundances and morphologies may hold important clues to the evolution of the IGM/ICM in these overdense regions. \citet{zirm09} showed that the halos around radio galaxies evolve strongly with redshift from $z\sim4$ to $z\sim1$, even though the AGN power remained the same. This indicates that it is primarily the surrounding medium that is evolving. As massive galaxies and their halos grow, a virial shock will create a quasi-static atmosphere of hot gas that may limit cooling and switch off pre-existing cold flows. We therefore expect to see a decrease in the luminosity and spatial extent of the extended \lya\ emission, eventually disappearing altogether, as protoclusters transition into clusters (Fig. \ref{fig:dekel_trends}).

It is currently still difficult to give a very precise answer to the question of whether protoclusters are preferentially found near radio galaxies or quasars and, if so, why (Sect. \ref{sec:rgqso}). Also, we must realize that the literature is likely to be biased, especially against the reporting of negative results. It is probably far more common for authors to publish new candidate protoclusters and spectroscopic confirmation of protoclusters than it is to publish non-detections. Although it is currently next to impossible to quantify its effect, this author suspects that for every positive detection there is likely to be at least one unpublished non-detection. This problem mainly affects the biased tracer studies, and is probably less of an issue in the large field-type surveys. We will also have to try to deal with the observational incompleteness. The fact that a particular observation produced no evidence of a protocluster around some target, does not mean that it is not present. \citet{venemans02} have shown that the total number abundance of radio galaxies, corrected for the typical radio synchrotron lifetimes, is consistent with every present-day BCG having gone through a radio galaxy phase at high redshift \citep[see also][]{miley08,hatch14}. The fact that there is strong evidence for protoclusters around high redshift radio galaxies, combined with the fact that their hosts are good candidates of (proto)BCGs supports this conclusion. The fact that the synchrotron lifetimes are relatively short also suggests that, for every protocluster traced by a radio galaxy (or radio-loud QSO), there should be at least several protoclusters without a radio source at the same epochs. This appears to be supported, at least qualitatively, by the fact that the number density of radio-quiet protoclusters is much higher than that of radio-loud protoclusters \citep[e.g.,][]{steidel05,chiang14,dole15,toshikawa16}, although a quantitative comparison must await future data from large blind surveys. It will also be interesting to see a definitive conclusion about the extreme (or not) environments of the luminous quasars at $z\sim6$ (Sect. \ref{sec:z6qsos}). Deep, wide, narrow-band surveys are most efficient in a number of atmospheric windows where the OH night-sky emission is low. \citet{banados13} therefore suggest to specifically search for QSOs at $z\approx5.7$, $z\approx6.6$ and $z\approx7.0$, and then target their environments in these narrow redshift intervals. For a complete census of their environments, these data should be complemented with ALMA in order to assess the possible contribution from obscured star-forming galaxies. At even higher redshifts, protoclusters may play an important role during the EoR (Sect. \ref{sec:reionization}).  

Until recently, the first stages of cluster evolution could only be explored indirectly, either through the fossil record in galaxy clusters or in cosmological simulations. However, starting with the first discoveries of protoclusters at $z\sim3$ in the late 1990s \citep[e.g.,][]{pascarelle96,lefevre96,steidel98}, we are finally beginning to catch a glimpse of this important epoch of cluster evolution. Although the existing observational sample is still relatively limited, this situation is rapidly changing owing to the next generation of wide and deep surveys that will ensure very large, uniform samples of protoclusters to be discovered (Sect. \ref{sec:surveys}). In the next few years, we can expect to see the boundary between the studies of ``clusters'' and ``protoclusters'' erode even further, eventually disappearing altogether, as we reach deeper into the realm of the protoclusters.  

\begin{acknowledgements}
I owe great gratitude to Arjun Dey, Avishai Dekel, Alessandro Rettura, Alvaro Orsi, Benedetta Ciardi, Brian Lemaux, Eduardo Ba{\~n}ados, Elizabeth Cooke, John Silverman, Jun Toshikawa, Kyoung-Soo Lee, Nina Hatch, Nobunari Kashikawa, Olga Cucciati, Ricardo Demarco, Sebastiano Cantalupo, Yi-Kuan Chiang, the editors, and the anonymous referees for valuable feedback that greatly improved this manuscript. I thank Lindsey Bleem for making available the cluster data used in Figs. \ref{fig:pcredshifts}, \ref{fig:pcmasses} and \ref{fig:mydekel}, and Mike Boylan-Kolchin, Benedetta Ciardi, Jun Toshikawa, Casey Stark, Yi-Kuan Chiang, Alexandro Saro, Sebastiano Cantalupo and Yusei Koyama for granting permission to reproduce their figures.
\end{acknowledgements}

% BibTeX users please use one of
\bibliographystyle{spbasic}      % basic style, author-year citations
%\bibliographystyle{spmpsci}      % mathematics and physical sciences
%\bibliographystyle{spphys}       % APS-like style for physics
%\bibliography{}   % name your BibTeX data base

% Non-BibTeX users please use

\end{document}